\documentclass[11pt]{article}
\usepackage{amssymb,dsfont,amsmath,amsfonts,hyperref}
\usepackage[active]{srcltx}
\usepackage{slashed}
\usepackage{color}
\advance \textheight by \topskip
\let\oldsqrt\sqrt
\def\sqrt{\mathpalette\DHLhksqrt}
\def\DHLhksqrt#1#2{%
\setbox0=\hbox{$#1\oldsqrt{#2\,}$}\dimen0=\ht0
\advance\dimen0-0.2\ht0
\setbox2=\hbox{\vrule height\ht0 depth -\dimen0}%
{\box0\lower0.4pt\box2}}

\def\be{\begin{equation}}
\def\ee{\end{equation}}
\def\w{\wedge}

\def\a{\alpha}
\def\b{\beta}

\def\e{\epsilon}
\def\l{\lambda}
\def\Comp{\mathbb{C}}
\def\Real{\mathbb{R}}
\def\rhoF{\rho_{\tiny {\cal F}}}

\begin{document}

\begin{center}
{\Large\bf Three-dimensional fractional-spin gravity}\\
\vspace*{10mm}
Nicolas Boulanger\footnote{\texttt{nicolasboul@gmail.com}; Associate
researcher of the FNRS, Belgium.}\\
\textit{Service de M\'ecanique et Gravitation,
Universit\'e de Mons - UMONS, Belgique}\\
\vspace*{5mm}
Per ~Sundell\footnote{\texttt{per.anders.sundell@gmail.com}}\\
\textit{Departamento de Ciencias F\'isicas, Universidad Nacional Andr\'es Bello - UNAB, Santiago, Chile}\\
\vspace*{5mm}
Mauricio ~ Valenzuela
\footnote{\texttt{valenzuela.u@gmail.com}}\\
\textit{Instituto de Ciencias F\'isicas y Matem\'aticas, Universidad Austral de Chile - UACH, Valdivia, Chile}\\ \textit{and} \\
\textit{Service de M\'ecanique et Gravitation, Universit\'e de Mons - UMONS, Belgique}
\end{center}

\vspace{1cm}

\begin{minipage}{.90\textwidth}

\paragraph{Abstract}
Using Wigner-deformed Heisenberg oscillators,
we construct 3D Chern--Simons models consisting of
fractional-spin fields coupled 
to higher-spin gravity and internal non-abelian gauge fields.
The gauge algebras consist of Lorentz-tensorial
Blencowe–-Vasiliev higher-spin algebras and compact
internal algebras intertwined by infinite-dimensional
generators in lowest-weight representations of the
Lorentz algebra with fractional spin.
In integer or half-integer non-unitary cases,
there exist truncations to $gl(\ell,\ell\pm 1)$ or
$gl(\ell|\ell\pm 1)$ models.
In all non-unitary cases, the internal gauge
fields can be set to zero.
At the semi-classical level, the fractional-spin 
fields are either Grassmann even or odd.
The action requires the
enveloping-algebra representation of the
deformed oscillators, while their
Fock-space representation suffices on-shell.
\\[5pt]
{\small
\textit{The project was funded in part by F.R.S.-FNRS ``Ulysse'' Incentive Grant for Mobility in Scientific Research.}}
\end{minipage}

\renewcommand{\thefootnote}{\arabic{footnote}}

\setcounter{footnote}{0}

\newpage

{\small \tableofcontents }

\numberwithin{equation}{section}

\section{Introduction}

\subsection{Anyons and statistical gauge fields}

An interesting feature of quantum physics in
three-dimensional spacetime is the presence of
identical particles with exotic statistics.
The basic notion dates all the way back to Leinaas
and Myrheim \cite{Leinaas:1977fm}
and later Wilczek \cite{Wilczek:1982wy}, who provided
specific models realizing such particles, which he
referred to as Anyons, as flux-charge quanta.
Subsequently, quantum field theories in flat spacetime
containing Anyons were constructed in \cite{Semenoff:1988jr,Frohlich:1988di,Forte:1990vb};
see also \cite{Forte:1990hd} for an early review.
Later, a group-theory approach based on wave-functions was developed in\footnote{Note also that equations of motion for massless,
fractional-spin particles in four-dimensional Minkowski spacetime have
been studied in \cite{Volkov1989,Klishevich:2001gy}. Although this formalism is covariant under infinitesimal Lorentz transformations, the four-dimensional Poincar\'e symmetry is violated for finite transformations \cite{Klishevich:2001gy}. } \cite{Jackiw:1990ka,Plyushchay:1991qd,Cortes:1992fa,Plyushchay:1997ie}; see \cite{oai:arXiv.org:1001.0274} for a  summary and extensions of these models.
More recently, the flux-charge realization has been
generalized to models with non-Abelian gauge fields
\cite{Itzhaki:2002rc}.
Anyons can also be realized without integrating out any statistical gauge fields either as clusters of
non-relativistic particles in two spatial dimensions kept together by external one-body potentials, such as a
simple harmonic potential, and interacting with each other only via boundary conditions imposed on the
multi-body wave functions \cite{Engquist:2008mc}, or as vertex operators in two-dimensional conformal
field theories \cite{Jatkar:1990re}.
For a more recent axiomatic treatise without gauge fields, see \cite{Mund:2008hn}.

The key idea is that at fixed time the configuration
space of a collection of massive particles whose trajectories
cannot coincide as the result of their interactions has
a non-trivial first homotopy group that is represented
non-trivially on the multi-body wave-functions or correlation
functions involving point-like operators \cite{Mund:2008hn}.
These representations thus furnish representations of
the braid group, which is why Anyon statistics is
synonymous to braid statistics.
The wave functions transform
under rotations with phase factors which can be
identified with the statistical phases under exchange of
identical particles.
Hence, one and the same phase characterizes
the statistics of the particles as well as
the representation of the spatial rotation group,
which is the essence of the generalized spin-statistics
theorem for massive particles in three-dimensional Minkowski
space with exotic statistics and fractional spin\footnote{In the case
of massless particles in three-dimensional Minkowski spacetime,
for which there does not exist any notion of helicity,
the statistics has been shown to instead be correlated
directly to the Lorentz spin in the case of
bosons and fermions \cite{Deser:1991mw}.
To our best understanding, so far there does not exist
any generalization of this result to fractional spins.
} \cite{Mund:2008hn}.
Thus, in 2+1 dimensions, the spin of a massive particle can
be an arbitrary real number, thereby providing and
interpolation between bosons and fermions.

In the realization of anyons in quantum field theories,
their fractional quantum numbers are typically quantum
effects due to the presence of Chern-Simons fields,
usually referred to as statistical gauge fields.
Their realizations as charged vortices \cite{Wilczek:1982wy}
and Hopf-interacting massive particles arise in effective
descriptions of matter-coupled Abelian Chern--Simons systems
\cite{Semenoff:1988jr}.
Integrating out the statistical Chern--Simons gauge field
produces effective topological non-local
Hopf interactions among the matter fields
that transmute their statistics; see also
\cite{Polyakov:1988md,Forte:1990vb,Forte:1990hd}
and \cite{Wu:1984kd,Wen:1988uw,Govindarajan:1992dr}
for related works using the $CP1$ formalism.
As for non-Abelian generalizations, the
conformal Chern-Simons-scalar \cite{Aharony:2012nh} and
Chern-Simons-fermion \cite{Giombi:2011kc} vector models
exhibiting level-rank type dualities providing examples
of three-dimensional Bose-Fermi transmutation \cite{Polyakov:1988md}.
In \cite{Aharony:2012nh} it is suggested that
these models contain Anyons at finite couplings.
Moreover, as proposed by Itzhaki \cite{Itzhaki:2002rc},
the statistical gauge fields can be taken to be
non-minimally coupled Yang-Mills fields by using Wilson lines for
connections shifted by the Hodge dual of the field strength to
generate the flux-charge bound states.

\subsection{Coupling of anyons to background fields}

On general grounds, one may ask whether Anyons can be described by any quantum-effective field theory that facilitates their coupling to ordinary tensorial and tensor-spinorial particles and fields, including gravity.
In an arbitrary curved background the description of Anyons
requires the introduction of a Lorentz connection valued in  non-(half-)integer spin representations of the Lorentz algebra, which  are infinite dimensional.
As such representations admit oscillator realizations, it seems
natural to incorporate them into Vasiliev's general
framework for higher-spin gravity \cite{Vasiliev:1992av,
Vasiliev:1999ba,Vasiliev:2003ev}.
The aim of this paper is to take a first step\footnote{See also the conference proceeding
\cite{Boulanger:2013zla}.} in this direction.

Vasiliev's equations provide a fully non-linear and
background-independent description
of a large class of higher-spin gravities in various dimensions,
including models with internal symmetry algebras
\cite{Konstein:1989ij,Prokushkin:1998bq} and
fermions
\cite{Konshtein:1988yg,Konstein:1989ij,Prokushkin:1998bq,
Sezgin:1998gg,Sezgin:2001yf,Vasiliev:2004cm},
of which some exhibit standard spacetime supersymmetry; for a recent review
in the case of four-dimensional higher-spin gravities, see \cite{Sezgin:2012ag}.
As far as spin-statistics relations are concerned, with notable exceptions in the presence of a positive cosmological constant \cite{Vasiliev:1986ye,Vasiliev:1986td,Sezgin:2012ag}\footnote{As observed by Vasiliev, in Lorentzian signature and in the presence of a positive cosmological constant, supergravities \cite{Vasiliev:1986ye} and linearized higher-spin supergravities \cite{Vasiliev:1986td} admit twisted reality conditions compatible with $\mathbb Z_2\times \mathbb Z_2$ graded quantum algebras; for a recent review and the extension to fully non-linear $dS_4$ higher-spin supergravities, see \cite{Sezgin:2012ag}. } or in Kleinian spacetime signature \cite{Sezgin:2012ag}, Vasiliev's higher-spin gravities have so far been assumed to consist of fields that are either bosonic Lorentz tensors or fermionic tensor-spinors \footnote{To our best understanding, this assumption on spin and statistics is required
for consistency only within the context of relativistic quantum field theories
in flat spacetimes of dimension four or higher;
see \emph{e.g.} \cite{Weinberg:1996kr}.
 Extensions to curved backgrounds of the
spin-statistics correspondence are given in \cite{Verch:2001bv} and
references therein.}.

However, our key observation is: Vasiliev's higher-spin gravities are not formulated \emph{a priori} in terms of Lorentz tensors and tensor-spinors; rather they are formulated in terms of master fields living on products of
space-time and fiber manifolds.
The latter contain non-commutative twistor or twistor-like spaces
whose coordinates generate the higher-spin and internal symmetry algebras.
The full specification of a Vasiliev-type higher-spin gravity model
thus requires the choice of a set of fiber functions that form an
associative algebra. Hence,
the incorporation of fractional-spin fields into the higher-spin framework
can be reduced to the technical problem of in which ways Vasiliev's theory
admits non-standard embeddings of the Lorentz connection leading to
fractional-spin representations.

The aim of this paper is to demonstrate within a simple class of models,
namely topological models of Chern--Simons type which we refer to as
fractional-spin gravities,
how standard tensorial higher-spin gravities can be extended by
fractional-spin fields by including additional sets of fiber functions that form
Lorentz representations characterized by arbitrary real-valued Lorentz spins.
As we shall see, the fractional spin fields appear within a bi-module of
one-forms acted upon by one-sides actions of the higher-spin algebra and
an internal color gauge group of infinite rank.
In doing so, a particular set of technical problems that we shall have to
address concerns the nature of infinite-dimensional representations and how
it is affected by different choices of bases.
To this end, we are going to focus on the on-shell formulation
of a class of Blencowe--Vasiliev
models \cite{Blencowe:1988gj,Bergshoeff:1989ns}
that arise within the Prokushkin--Vasiliev
system \cite{Prokushkin:1998bq} as a consistent truncation.

\subsection{Outline of the paper}

In Section 2, which can be skipped at first reading, we collect further
background material, general remarks on higher-spin gravities in three
dimensions and how our fractional-spin gravity models can be embedded
into this context.
We then summarize our main results, including material from a work in progress.
In Section 3, we then proceed with the main analysis of anyon
representations in $AdS_3$ and their realizations using the
Wigner-deformed Heisenberg oscillators.
In this section we shall stress details concerning the infinite-dimensional nature of these representations and in particular the importance of keeping track of their indecomposable structures in critical limits and related choices of bases will be stressed in Section \ref{Sec:bases}.
In Section 4, the fractional-spin Chern--Simons theory is formulated and some of its truncations are presented.
We conclude in Section 5.

\section{Preliminary  remarks and summary}

In this section we review some features of higher-spin gravity that are of conceptual interest and of importance for generalizations of our models.
We then summarize our results including some material concerning mainly
the off-shell formulation to be presented elsewhere.
As the contents of this section are not crucial for the main
analysis in the coming sections, to which the reader may 
therefore skip immediately if so desired.

\subsection{Preliminary remarks on three-dimensional higher-spin gravities}

\paragraph{Three-dimensional higher-spin gravity landscape.}
Three-dimensional topological higher-spin gravities with
Lorentz-tensorial and tensor-spinorial gauge fields
are described semi-classically by the Fradkin--Vasiliev-inspired Blencowe actions \cite{Blencowe:1988gj}.
These theories are of Chern--Simons type and
based on Lie algebras generated by ordinary Heisenberg oscillators, or
equivalently, area preserving diffeomorphisms of two-spheres and
two-hyperboloids \cite{Bergshoeff:1989ns}.
As pointed out by Vasiliev \cite{Vasiliev:1989re}, these algebras admit
deformations based on Wigner-deformed Heisenberg oscillators
\cite{Wigner:50,Yang:51}, or equivalently, algebras of
symplectomorphisms of fuzzy two-hyperboloids and two-spheres.

These topological models sit inside a
larger landscape of matter-coupled higher-spin
gravities described by the Prokushkin--Vasiliev
equations \cite{Vasiliev:1996hn,Prokushkin:1998bq};
see also \cite{Barabanshchikov:1996mc}.
Although their structure resembles that of the higher-dimensional Vasiliev
equations \cite{Vasiliev:1990en,Vasiliev:2003ev},
the three-dimensional higher-spin gravities exhibits a proper feature:
its dynamical Weyl zero-forms are necessarily accompanied by
topological Weyl zero-forms\footnote{As pointed out to us by D. Jatkar,
it is natural to think of these topological degrees of freedom in 
higher-spin gravity
as corresponding two-dimensional conformal field theory defects.} while the
corresponding sectors can be consistently
truncated in four and higher dimensions.

In any dimension, there exists a special topological zero-form
(which is a singlet) that can acquire an expectation value,
$\nu$ say, that deforms the higher-spin symmetries.
However, it is only in three dimensions that $\nu$ does not deform
the anti-de Sitter vacuum.\footnote{In four dimensions,
the maximal finite sub-algebra
of the higher-spin algebra that is preserved by $\nu$ is
$so(1,3) $ or $so(2,2)$ depending on the choice of signature.
This suggests that four-dimensional fractional-spin gravities
correspond holographically to three-dimensional massive quantum
field theories with anyons, and that
these models are integrable in a suitable sense,
as the higher-spin symmetries are deformed rather than broken.}
The expansion around this $AdS_3$-vacuum, with its expectation
value $\nu$, yields the aforementioned Chern--Simons models based on
deformed oscillators as consistent truncations (upon setting all fluctuations
in the Weyl zero-form to zero).
In particular, for critical values of $\nu\,$, given conventionally by
$\nu=-2\ell-1$ with $\ell=0,1,2,\dots\,$,
the higher-spin algebras contain $gl(2\ell+1)$
subalgebras \cite{Vasiliev:1989re},
and the Chern--Simons models can be reduced further
down to $sl(N|N\pm 1)$ and pure bosonic $sl(N)$
models studied in \cite{Bergshoeff:1989ns}.

\paragraph{Prokushkin--Vasiliev system and Wigner-deformed
Heisenberg oscillators.}

The Prokushkin--Vasiliev system consists of a connection one-form
$\widehat A$ and matter zero-form $\widehat B$ living on a base manifold given
locally by the direct product of a commutative spacetime ${\cal M}$
and non-commutative twistor space ${\cal Z}$ with a closed and central two-form $\widehat J\,$.
These master fields are valued in associative algebras consisting of functions
on a fiber manifold ${\cal Y}\times {\cal I}\,$,
the product of an additional twistor space ${\cal Y}$
and an internal manifold ${\cal I}$ whose coordinates
generate a matrix algebra.
The Prokushkin--Vasiliev field equations, \emph{viz.}
$\widehat {\rm d}\widehat A+\widehat A^2+\widehat J \widehat B=0$ and
$\widehat {\rm d}\widehat B+[\widehat A,\widehat B]=0$,
state that $\widehat A=\widehat A|_{\cal M}+\widehat A|_{\cal Z}$
describes a flat connection on ${\cal M}$ and a pair of oscillators on
${\cal Z}\times {\cal Y}$ deformed by local as well as topological degrees
of freedom contained in $\widehat B\,$.

Working within the fully non-linear system
its constructors observed that models with sufficiently elaborate internal
algebra admit $AdS_3$-vacuum expectation
values  \cite{Prokushkin:1998bq}
\begin{eqnarray}
\langle \widehat B \rangle =\nu\ ,
\end{eqnarray}
and that the perturbative expansions around these vacua yield
parity-invariant three-dimensional higher-spin gravities containing massive scalars.\footnote{These scalars behave as massive higher-spin fields for critical values of $\nu$; whether parity invariance can be broken within the Prokushkin--Vasiliev formalism remains an open issue.}
After a suitable redefinition, the perturbatively-defined master fields
become valued in associative algebras
\be {\cal A}(2;\nu;{\cal I})=\bigoplus_{\Sigma} {\cal A}_{\Sigma}\ ,\ee
where ${\cal I}$ refers to sets of internal generators (including the $\mathbb Z_2$-generator $\Gamma$ used to double $sl(2)$ to $sl(2)_+ \oplus sl(2)_-$),
consisting of sectors ${\cal A}_\Sigma$ of suitable non-polynomial
extensions of the universal enveloping algebra $Aq(2;\nu)$
\cite{Vasiliev:1989re} of the Wigner-deformed Heisenberg oscillator algebra
\cite{Wigner:50,Vasiliev:1989re,Plyushchay:1997mx}\footnote{Blencowe's
construction \cite{Blencowe:1988gj} makes use of the undeformed
algebra $Aq(2;0)_+ \oplus Aq(2;0)_-\,$.} ($\alpha=1,2$)
\begin{equation}
[q_\alpha,q_\beta]=2i\epsilon_{\alpha\beta}(1+\nu k)\ ,\qquad
\{k,q_\alpha \}=0\ ,\qquad k^2=1\ .
\label{defy}
\end{equation}
Thus, as the fully non-linear formulation rests on associative differential
algebras,
one may ask whether these can be extended by adding sectors of
composite operators and refining correspondingly the star-product
composition rule as to retain associativity, thus allowing the formal structure
of the full master-field equations to remain intact,
with the aim of facilitating modified embeddings of the Lorentz algebra
into the gauge algebra that produces perturbatively-defined field contents
containing fractional-spin fields.

Indeed, as we shall outline next, this can be done in a relatively straightforward fashion by
adding sectors of non-polynomial operators corresponding
to Fock-space endomorphisms.
These operators are given essentially
by star-product versions of vacuum-to-vacuum projectors
dressed by left and right multiplications
by arbitrary polynomials.
The extended associative star-product rules can then be defined using
a matrix structure.

\subsection{Outline of fractional-spin gravities}\label{Sec:main}

\paragraph{Matrix fusion rules\,.} The fractional-spin gravities that we shall consider
are based on $\mathbb Z_2$-graded associative algebras that are formed by extending the enveloping algebra $Aq(2;\nu)$ by sectors of operators acting in the Fock space ${\cal F}$ consisting of states with distinct eigenvalues of the spatial spin generator $J_0$. More formally, we define
\be {\cal A}(2;\nu|\mathfrak{o}(2)_{J_0};{\cal F}):=\left[\begin{array}{cc} \overline{Aq}(2;\nu)_{++} &
\rhoF({\rm End}({\cal F}))_{+-}\\ \rhoF({\rm End}({\cal F}))_{-+}&
\rhoF({\rm End}({\cal F}))_{--}\end{array}\right]
\ee
where the injective homomorphism, or monomorphism,
\begin{eqnarray}\label{rhomap}
\rhoF \quad : \quad {\rm End}({\cal F})\quad  \hookrightarrow \quad \overline{Aq}(2;\nu)
\end{eqnarray}
maps the space
\be {\rm End}({\cal F}):=\left\{\check E:=\sum_{m,n\geqslant 0} E^{mn}|m\rangle\langle n|\right\}\cong {\rm Mat}_{\infty}(\Comp)\ee
of endomorphisms of the Fock space
\be {\cal F}=\sum_{m=0}^\infty \Comp\otimes |m\rangle\ ,\quad (\check N-m)|m\rangle =0\ ,\ee
of the undeformed Heisenberg oscillator algebra
$[\check b^-,\check b^+]=1$ with number operator
\begin{eqnarray}
\check N := {\check b}^+ \,{\check b}^-
\end{eqnarray}
into a non-polynomial completion
\begin{equation}\label{fqk}
\overline{Aq}(2;\nu)\,:=\left\{\ f(k;q)=\sum_{m=0,1; n\geq 0}  k^m{}
f_{m;(n)}(q)\ ,\quad f_{m;(n)}(q):=f_{m;(n)}^{\,\alpha_1 \cdots \alpha_n} {}
q_{(\alpha_1 } {}\cdots{} q_{\alpha_n)} \,\right\}\,,
\end{equation}
of the enveloping algebra $Aq(2;\nu)$ of the Wigner-deformed Heisenberg oscillator algebra \eqref{defy}. The monomorphism $\rhoF$ is defined by the rule
\be \label{rhoF}
\rhoF(|m\rangle\langle n|)=P_{m|n}\ ,\ee
where the generalized projectors $P_{m|n} \in \overline{Aq}(2;\nu)$ obey
\be
P_{m|n}{} P_{k|l}=\delta_{nk} P_{m|l}\ ,\quad (N_\nu-m){} P_{m|n}=0=P_{m|n}{}(N_\nu-n)\ ,
\ee
with number operator $N_\nu:=\rhoF(\check N)$ related to
$J_0$ by
\begin{eqnarray}
\label{J01} N_\nu := 2 J^0 - \tfrac12 (1+\nu) \; ,
\quad
J^0   :=  \tfrac{1}{4}\, \{ a^- , a^+\}\; ,
\end{eqnarray}
expressed in terms of the \emph{deformed} oscillators $(a^-,a^+)$ obeying
$[a^- , a^+] = 1 + k\,\nu\,$.
The relationship \cite{Plyushchay:1997mx} between deformed and undeformed oscillators $(a^- , a^+)$ and $(b^- , b^+)$, respectively, is presented in Subsection \ref{subsec:RepresentationAq}.
In defining ${\cal A}(2;\nu|\mathfrak{o}(2)_{J^0};{\cal F})$ we have also used
\be\label{projector} \overline{Aq}(2;\nu)_{\sigma\sigma'}
:=\Pi_{\sigma}\,{}\,\overline{Aq}(2;\nu)\,{}\,\Pi_{\sigma'}\ ,\quad \Pi_\pm=\frac12(1\pm k)\ .
\ee
The associative composition rule in ${\cal A}(2;\nu|\mathfrak{o}(2)_{J^0};{\cal F})$
is defined as follows: Let
\be \mathbb{M}_i=\left[\begin{array}{cc} A_i& \rhoF(\check B_i)\\\rhoF(\check{C}_i) & \rhoF(\check{D}_i)\end{array}\right]\ , \quad i=1,2\ ,\ee
be two elements in ${\cal A}(2;\nu|\mathfrak{o}(2)_{J^0};{\cal F})$ with $A_i\in Aq(2;\nu)$ being finite polynomials and $\check B_i,\,\check C_i,\, \check{D_i} \in {\rm Mat}_{K}(\Comp)\subset {\rm End}({\cal F})$ being finite matrices,
that is, $\check B_i=\sum_{m,n=0}^K A^{m,n}_i|m\rangle\langle n|$ \emph{idem}
$\check C_i$ and $\check D_i$ ($i=1,2$).
Then $\mathbb{M}_1 \mathbb{M}_2$ is defined by standard matrix multiplication followed by star-product compositions and expansions of the results in the appropriate bases.
In particular, the quantity $\rhoF(\check B_1) \rhoF(\check C_2) = \sum_{m,n,p=0}^K  M_1^{mp} M_2^{pn} P_{m|n}$ is given by its expansion in the monomial basis of $\overline{Aq}(2;\nu)$, while $A_1 \rhoF(\check B_2)$ and $\rhoF(\check C_1) A_2$ are expanded in the matrix basis of ${\rm End}({\cal F})$.
This composition rules is then extended to ${\rm End}({\cal F})$ by allowing the degree of $A_i$ and $K$ to be arbitrarily large.
Thus, the fractional-spin algebra has a product rule that combines star-product compositions in initial bases followed by expansions of the results in a
final bases,
which on may refer to as a \emph{fusion rule}.
We note that in the case at hand, the fusion rule does not require
any expansion of $P\in Aq(2;\nu)$ in the matrix basis of ${\rm End}({\cal F})\,$.

\paragraph{Hermitian conjugation\,.}

Defining the hermitian conjugation operation $\dagger$ in $Aq(2;\nu)$ by
\be (q_\a)^\dagger=q_\a\ ,\quad (k)^\dagger=k\ ,\ee
it follows that the Fock-space realization $\check q_\a$ of the
deformed oscillators \cite{Plyushchay:1997mx,Plyushchay:1997ty} with
\begin{equation}
\check k=\varepsilon (-1)^{\check N}=\varepsilon \cos (\pi \check N)\,,\quad
(\check N-n)|n\rangle=0\ ,\quad \varepsilon^2=1\ ,
\end{equation}
obeys
\be (\check q_\a)^{\check{\dagger}}=\check C \check q_\a \check C\ ,\ee
where $\check{\dagger}$ refers to the standard hermitian conjugation operation in ${\rm End}({\cal F})$ and the charge conjugation matrix $\check C$ is given by the identity in the unitary regime
$\varepsilon \nu\geqslant -1$ and a non-trivial matrix in the non-unitary regime $\varepsilon \nu<-1$.
Assuming furthermore that $\check C^2=1$ and $(\check C)^{\check{\dagger}}=\check C\,$,
it follows that
\be
\dagger \circ \, \rhoF=\rhoF \circ {\rm Ad}_{\check C}\circ \check{\dagger}={\rm Ad}_{C}\circ \rhoF\circ \check{\dagger}\ ,\quad C:=\rhoF(\check C)\ ,\ee
or more explicitly, if $f=\rhoF(\check f)$ then
\be f^\dagger=\rhoF(\check C\check f^{\check{\dagger}}\check C)=C{} \rhoF(\check f^{\check{\dagger}}){} C\ .
\ee

\paragraph{Master gauge fields\,.}

Starting from a Prokushkin--Vasiliev model with $\langle \widehat B\rangle =\nu$ and
fiber algebra
\be
{\cal A}_\sigma :=\left[{\cal A}(2;\nu|\mathfrak{o}(2)_{J^0};{\cal F})\otimes {\rm Cliff}_1(\Gamma)\otimes {\rm Cliff}_1(\xi)\right]_\sigma\ ,
\quad \sigma = \pm\ ,
\ee
where ${\rm Cliff}_1(\Gamma)$ and ${\rm Cliff}_1(\xi)$ denote, respectively,
a bosonic Clifford algebra and a fermionic
Clifford algebra with respective generators obeying
\be
\Gamma{} \Gamma=1=\xi{} \xi\ ,\quad \e_{\rm s}(\Gamma,\xi)=(0,1)\
\ee
and where $\e_{\rm s}$ denotes the Grassmann parity,
we may consider the consistent truncation $\widehat B=\nu$,
leaving the flat connection
\begin{equation}
\mathbb{A}_\sigma =\left[
\begin{array}{cc}
W & \psi_\sigma \\
\overline{\psi}_\sigma  & U
\end{array}
\right]\quad \in \quad \Omega^{[1]}({\cal M}_3)\otimes {\cal A}_\sigma \ ,
\qquad \sigma = \pm\ .
\end{equation}
We demand the master fields to be Grassmann-even, \emph{i.e.}
\be \e_{\rm s}(W,\psi_\pm,\overline{\psi}_\pm,U)=(0,0,0,0)\ ,\ee
and to have intrinsic parities
\be \sigma(W,\psi_\pm,\overline{\psi}_\pm,U)=(+1,\pm 1,\pm 1,+1)\ ,\ee
where $\sigma$ is defined on polynomials $f(q,k,\xi)$ of definite degrees in
$q^{\alpha}$ and $\xi$ by
\be
\pi_q \pi_\xi(f) =: \sigma(f) f\ ,\label{sigmadef}\ee
where we used the automorphisms
\be \pi_q(f(q,k,\xi)):=f(-q,k,\xi)\ ,\quad \pi_{\xi} f(q,k,\xi):=f(q,k,-\xi)\ .\ee
Taking into account the $\Pi_\pm$-projections and assigning the following Grassmann parity
\begin{equation}
\label{statisticsq}
\epsilon_s(q_{\alpha}) = 0 \,
\end{equation}
it follows that $(W,U)$ and $(\psi_-,\overline{\psi}_-)$ are $\xi$-independent, hence consisting of
Grassmann-even component fields, while $(\psi_+,\overline{\psi}_+)$ are linear in $\xi$, hence
consisting of Grassmann-odd component fields.
We note that $\psi_\pm$ and $\overline{\psi}_\pm$, respectively, transform under the left actions of $\overline{Aq}(2;\nu)_{++}\otimes {\rm Cliff}_1(\Gamma)$ and
$\rhoF({\rm End}({\cal F}))_{--}\otimes {\rm Cliff}_1(\Gamma)$ and under the right actions of
$\rhoF({\rm End}({\cal F}))_{--}\otimes {\rm Cliff}_1(\Gamma)$ and
$\overline{Aq}(2;\nu)_{++}\otimes {\rm Cliff}_1(\Gamma)$.

The reality conditions on $\mathbb{A}_\sigma$ will be
chosen such that $W$ belongs to a non-compact real form of
$\overline{Aq}(2;\nu)_{++}  \otimes {\rm Cliff}_1(\Gamma)$
containing the Lorentz generators $\Lambda^{\a\b}\Pi_+{} q_{(\a}{} q_{\b)}{} \Pi_+$
with $(\Lambda^{\a\b})^\dagger=\Lambda^{\a\b}\,$,
while
$U\in u(\infty) \otimes {\rm Cliff}_1(\Gamma)\,$.
We note that for generic $\nu$, the model may be level truncated such that
\begin{equation} U\in u(K)\oplus u(K) \ ,\quad (\psi_\pm,\overline{\psi}_\pm)\in (K,\overline K)\ ,\end{equation}
for $K=1,2,\dots,\infty$, but that the more interesting truncations arise spontaneously as $\nu$ assumes critical values.

\paragraph{Embedding of Lorentz algebra\,.}

A standard space-time formulation of the Chern--Simons
field theory requires the choice of a canonical Lorentz connection $\omega \in sl(2)_{\rm Lor}$ associated with a principal Lorentz bundle over ${\cal M}_3$.
In general, the Lorentz algebra can be embedded into the gauge algebra in several inequivalent ways
leading to physically distinct models.
In particular, one has
\begin{itemize}
\item the diagonal embedding
\be sl(2)_{\rm Lor}=sl(2)_{\rm diag}:={\rm span} \left\{ q_{(\a} {} q_{\beta)}\right\}\ ,
\label{standardso12}
\ee
which yields standard higher-spin (super)gravities consisting of Lorentz tensors (and tensor spinors);
\item the alternative non-diagonal embedding
\be sl(2)_{\rm Lor}=\Pi_+ {} sl(2)_{\rm diag}=\left\{\Pi_+ {} q_{(\a} {} q_{\b)}\right\}\ ,\label{anyonso12}
\ee
which yields the fractional-spin (super)gravities in which the canonical Lorentz connection $\omega$ is thus embedded in $W$ such that $\psi$ and $\bar \psi$, respectively, transform in left- and right-modules with fractional Lorentz spin.
\end{itemize}

\paragraph{Supertrace and action\,.}

The non-polynomial completion $\overline{Aq}(2;\nu)$ of the enveloping algebra $Aq(2;\nu)$ admits the trace operation
\be\label{Trace} {\rm Tr}_\nu(f):= {\rm STr}_\nu(k{} f)\ ,\ee
where the supertrace operation ${\rm STr}_\nu$ is fixed uniquely by its
defining properties
\be {\rm STr}_{\nu}(f{} g)=(-1)^{\tfrac{1-\sigma(f)}{2}}{\rm STr}_{\nu}(g{} f)
= (-1)^{\tfrac{1-\sigma(g)}{2}}{\rm STr}_{\nu}(g{} f)
\ ,\quad {\rm STr}_{\nu}(1):=1\ ,\ee
where the intrinsic parity $\sigma$ is defined
in \eqref{sigmadef}, \emph{i.e.}
$f(-q,k)=\sigma(f) f(q,k)\,$.
Using the Weyl-ordered basis \eqref{fqk}, one has \cite{Vasiliev:1989re}
\be
\label{STr}{\rm STr}_\nu (f(q,k))= f_{0;(0)} - \nu \, f_{1;(0)}\ .\ee
Upon including the Clifford algebras, we define
\be
{\rm Tr}(f)={\rm Tr}_{\nu}(f)|_{\xi=0=\Gamma}\ ,\quad f\in \overline{Aq}(2;\nu)\otimes
{\rm Cliff}_1(\Gamma)\otimes {\rm Cliff}_1(\xi)\ ,
\ee
and equip ${\cal A}_\pm$ with the trace operation
\be {\rm Tr}_\mp (\mathbb{M}_\pm)={\rm Tr}(A\mp D)\ ,\quad
\mathbb{M}_\pm=\left[\begin{array}{cc} A&B_\pm\\
C_\pm&D\end{array}\right]\in{\cal A}_\pm\ ,\label{traceoperation}\ee
where thus $\sigma(B_\pm)=\sigma(C_\pm)=\pm 1$, which
obeys\footnote{Expanding the zero-form
$B_\mp=\sum_I B_\mp^{I} \Theta^\mp_I$ where
$\Theta^\mp_I$ denote a basis of composite operators and $B_\mp^I$
component zero-form fields with $\e_{\rm s}(\Theta^\mp_I)=\e_{\rm s}(B_\mp^I)=(1\pm 1)/2\,$ \emph{idem} for $C_\pm$,
it follows from ${\rm Tr}(\Theta^\mp_I{} \Theta^\mp_J)={\rm Tr}(\Theta^\mp_J{} \Theta^\mp_I)$
that ${\rm Tr}_\pm (\mathbb{M}_{1;\mp} {}\mathbb{M}_{2;\mp})={\rm Tr}(A_1{} A_2\pm D_1 {} D_2)\pm \sum_{I,J} (B_{1;\mp}^I C_{2;\mp}^{ J}+ B_{2;\mp}^I C_{1;\mp}^J){\rm Tr}(\Theta^\mp_I{} \Theta^\mp_J)$ is symmetric under the exchange $1 \longleftrightarrow2\,$.}
\be {\rm Tr}_\pm (\mathbb{M}_{1;\mp} {}\mathbb{M}_{2;\mp})= {\rm Tr}_\pm (\mathbb{M}_{2;\mp}{}\mathbb{M}_{1;\mp})\ .\ee
The Chern--Simons action
\begin{eqnarray}
S[\mathbb{A}_\pm]&=&\int \; {\rm Tr}_\mp
\left( \tfrac12\mathbb{A}_\pm{} {\rm d} \mathbb{A}_\pm+\tfrac{1}{3} (\mathbb{A}_\pm)^{{} 3} \right) \label{CS1}
\\
[5pt]&=&\int  \; {\rm STr}_\nu\left( \tfrac12 W {} {\rm d} W+\tfrac{1}{3} W^{{} 3}+W{} \psi_\pm{} \bar\psi_\pm\right.\\
 &&\left. \pm  ( \tfrac{1}2 U {} {\rm d} U+\tfrac{1}{3} U^{{} 3}+U{} \bar\psi_\pm {} \psi_\pm) + \tfrac12(\psi_\pm {}
 {\rm d} \bar{\psi}_\pm\pm \bar{\psi}_\pm {} {\rm d} \psi_\pm)\right)|_{\Gamma=0=\xi}\qquad \label{CS2}\ ,
\end{eqnarray}
as can be seem using the $\Pi_\pm$ projections and
\be {\rm STr}_{\nu}(\bar \psi_\pm {} W{} \psi_\pm)=\pm {\rm STr}_{\nu}(W{} \psi_\pm{} \bar \psi_\pm)\ ,\quad {\rm STr}_{\nu}(\psi_\pm {} U{} \bar\psi_\pm)=\pm {\rm STr}_{\nu}(U{} \bar\psi_\pm{} \psi_\pm)\ .\ee
On $\rhoF({\rm End}({\cal F}))$ the operation ${\rm STr}_\nu$ reduces to the standard Fock-space supertrace, \emph{viz.}
\be {\rm STr}_\nu(\rhoF(\check f))={\rm STr}_\nu\big(\rhoF(|0\rangle\langle 0|)\big) \sum_{m=0}^\infty (-1)^m \langle m|\check f|m\rangle\ .\ee
Thus, the level of the internal gauge algebra is proportional to the $\nu$-dependent quantity
${\rm STr}_\nu\big(\rhoF(|0\rangle\langle 0|)_{--}\big)\equiv {\rm STr}_\nu(\Pi_-{} P_{0|0})\,$. 

\paragraph{On-shell formulation in the Fock space\,.}

The equations of motion take the form
\begin{equation}\label{F}
{\mathbb{F}}{}_\pm:={\rm d} {\mathbb{A}}{}_\pm
   + ({\mathbb{A}}{}_\pm)^2=0 \,,
\end{equation}
that is,
\be
{\rm d}   W +  W^{{} 2} +
\psi_\pm{} {\overline{\psi}}{}_\pm=0\ ,\quad
{\rm d}  U +  U^{{} 2} + {\overline{\psi}}{}_\pm {} \psi_\pm = 0\ ,
\ee
\be {\rm d}  \psi_\pm + W {} \psi_\pm
+ \psi_\pm  U=0\ ,\quad
{\rm d} {\overline{\psi}}{}_\pm +
 U {} {\overline{\psi}}{}_\pm + {\overline{\psi}}{}_\pm {}  W=0\ .
\ee
Assuming that $W$ lies in the image of $\rhoF$, one can thus equivalently work on-shell with the
Fock-space presentation of the equations of motion, \emph{viz.}
\be
{\rm d} \check W+\check W^2+\check \psi_\pm\check{\overline{\psi}}{}_\pm=0\ ,\quad
{\rm d}\check U+\check U^2+\check{\overline{\psi}}{}_\pm \check \psi_\pm=0\ ,
\ee
\be
{\rm d} \check \psi_\pm+\check W\check \psi_\pm+\check \psi_\pm \check U=0\ ,\quad
{\rm d} \check{\overline{\psi}}{}_\pm +
\check U \check{\overline{\psi}}{}_\pm + \check{\overline{\psi}}{}_\pm\check W = 0\ ,
\ee
which we shall analyze in more detail below, though we note that the calculation of the action requires the star-product formalism.

\paragraph{Fractional Lorentz spin\,.}
By the construction outlined so far, and working in conventions where
\be sl(2)_{\rm diag}= \left\{ J^+, J^0, J^{-}\right\}=\left\{\tfrac{1}{2}\, a^+ {} a^+,\; \tfrac{1}{4}\,\{a^+,a^-\}_{}, \;\tfrac{1}{2}\,a^-{} a^-\right\}\ ,\ee
where the deformed ladder operators $a^\pm$ are linear combinations of $q_\a$ obeying
\be
[a^-,a^+]_{}=1+\nu k\ ,\qquad \{k,a^\pm\}_{}=0\ ,\qquad (a^\pm)^\dagger=a^\mp\ ,\label{dha}
\ee
the Lorentz spin of $(\psi,\bar\psi)$, say $\alpha$, defined to be the lowest weight of the generator
\be
J^0=\tfrac12 N_{\nu}+\tfrac14(1+\nu)\ ,
\label{J0}
\ee
is one of the roots of the quadratic Lorentz Casimir
\be
\label{casimir}
C_2(sl(2)_{\rm Lor}){} \psi_\pm=-\alpha(\alpha-1)\,\psi_\pm  \;,  \qquad \bar{\psi}_\pm{} C_2(sl(2)_{\rm Lor})  =-\alpha(\alpha-1)\,\bar{\psi}_\pm\,.
\ee
Taking into account $k=\varepsilon (-1)^{N_\nu}\,$, one has
\begin{equation}\label{alpha}
\alpha=\frac{1+\nu}4+\frac{1-\varepsilon}{4}\ ,
\qquad \varepsilon = \pm 1\ .\end{equation}
The Lorentz spin of $(\psi,\bar\psi)$ is thus fractional and hence $(\psi,\bar\psi)$ transform in an
infinite-dimensional irreducible representation of $sl(2)_{\rm Lor}$ except for critical values of $\nu$.
In the following, we will implicitly assume that $\varepsilon=+1\,$ unless explicitly mentioned
otherwise.

\paragraph{Critical $\nu$\,.}

For the
\be \mbox{critical values}:~ \nu~=~\nu_\ell~:=~-2\ell-1\ ,\quad \ell=0,1,2,\dots\ ,\ee
the deformed Wigner-Heisenberg algebra is known \cite{Vasiliev:1989re,Plyushchay:1997mx,Plyushchay:1997ty}
to admit $(2\ell+1)$-dimensional irreducible representations.
As we shall see in Section \ref{subsec:RepresentationAq},
the algebra $Aq(2;\nu)$ can be represented in number of
ways on $\mathcal F$, leading to representations $\check{Aq}(2;\nu;\tau)$
whose indecomposable structures for critical $\nu$ depend on
a parameter $\tau\in \mathbb{R}$.
In particular, it is
possible to choose the representation matrices in accordance
with the direct-sum decomposition
\be \check{Aq}(2; - 2\ell -1;0 ) \cong gl(2\ell+1)\oplus
\check{Aq}(2;2\ell+1;0)\ ,
\ee
where $\check{Aq}(2; 2\ell + 1)$ is isomorphic to the
representation of ${Aq}(2;-2\ell - 1)$ in ${\cal F}$
on the singular vector $\vert 2\ell+1\rangle$.
Thus, in critical limits, the indecomposable structures of
${Aq}(2;-2\ell-1)_{++}$ and ${Aq}(2; - 2\ell -1 )_{--}$ differ from
those of $\check{Aq}(2;-2\ell-1;\tau)_{++}$ and $\check{Aq}(2;- 2\ell -1;\tau )_{--}$, respectively, though in both cases the the finite-dimensional
sectors that remain after factoring out the ideals are isomorphic to
$gl(2\ell+\tfrac12(1+\varepsilon))$ and $gl(2\ell+\tfrac12(1-\varepsilon))$, respectively.

\paragraph{Generalized $\mathfrak h_{\rm 1-sided}$\,.}

Finally, the fractional-spin gravity admits a natural generalization based on the Fock space
${\cal F}$, in which $\check J^0$ is diagonal, and an additional state space
\be \tilde{\cal F}=\bigoplus_{\l}\Comp\otimes |\l\rangle\ ,\qquad (\check H-\l)|\l\rangle=0\ ,\ee
where $\check H$ is a Hamiltonian with normalizable (bound) states.
If there exists a ${}$-product implementation with fusion rules corresponding to
\be
W\in \overline{Aq}(2;\nu){}_{++}\ ,\qquad U\in \rho_{\tilde{\cal F}}({\rm End}(\tilde{\cal F}))\ ,
\qquad \psi\in \rho_{\tilde{\cal F}}({\rm End}(\tilde{\cal F}))\ ,\qquad \bar\psi\in\rho_{\tilde{\cal F}}({\rm End}(\tilde{\cal F}))\ ,
\ee
where $\rho_{\tilde{\cal F}}:{\rm End}(\tilde{\cal F})\rightarrow \overline{Aq}(2;\nu)$, then we propose a Chern--Simons action based on the Killing form
\be {\rm STr}\big(\rho_{\tilde{\cal F}}(\sum_{\l,\l'}|\l\rangle\langle \l'|f^{\l\l'})\big)=\sum_\lambda {\rm STr}(P_{\l|\l}) f^{\l\l}\ ,\ee
where $\sum_{\l,\l'}|\l\rangle\langle \l'|f^{\l\l'}\in{\rm End}(\tilde{\cal F})$ and $P_{\l|\l}$ is the star-product algebra element corresponding to
$|\lambda\rangle\langle\lambda|$.

\section{The Wigner-deformed Heisenberg oscillator algebra}
\label{sec:Wigner-deformed}

This section describes the concrete explicit realization of the fractional-spin algebras using
Wigner-deformed Heisenberg oscillators.

\subsection{The enveloping algebra $Aq(2;\nu)$ and its derived Lie (super)algebras }\label{Aqsec}

The universal enveloping algebra $Aq(2;\nu)$ of the Wigner-deformed Heisenberg oscillator algebra is
the associative algebra spanned by arbitrary polynomials in the deformed oscillators $q_\a$ and the
Kleinian $k$ modulo their relations \eqref{defy}.
It contains two associative subalgebras given by its subspaces
$Aq(2;\nu)_{\pm\pm}=\Pi_{\pm} Aq(2;\nu) \Pi_\pm$, where $\Pi_\pm=\frac12(1\pm k)\,$.
By taking $[f_1,f_2]:=f_1 f_2 -f_2 f_1$, these algebras turn into Lie algebras,
which we denote by $lq(2;\nu)$ and $lq_{\pm\pm}(2;\nu)$,
in their turn containing the Lie subalgebras $slq(2;\nu)$ and $slq(2;\nu)_{\pm\pm}\,$, respectively,
obtained by factoring out $\Comp\otimes \mathbf 1$ and $\Comp\otimes \Pi_\pm\,$.
The algebra $Aq(2;\nu)$ can also be endowed with the structure of a $\mathbb{Z}_2$-graded Lie algebra, denoted by $q(2;\nu)$, with graded commutator
\be
[f_1,f_2]_{\varepsilon}
:=f_1f_2-(-1)^{\varepsilon(f_1)\varepsilon(f_2)}f_2 f_1\ ,
\label{gradedcomm}
\ee
with degree defined by\footnote{We use (square) curved brackets to denote strength-one
(anti) symmetrization.}
\be
\varepsilon(k^B  \, q_{(\alpha_1 } \cdots\, q_{\alpha_n)}):= \tfrac12(1-(-1)^n)\ ,\ee
that is, $\varepsilon(f(q,k)) =0$ if $f(-q,k)=f(q,k)$ and $ \varepsilon(f(q,k)) =1$ if $f(-q,k)=-f(q,k)$.
Factoring out the identity from $q(2;\nu)$ yields a superalgebra, which we denote by $sq(2;\nu)$.
The Lie algebras $lq(2;\nu)$ and $slq(2;\nu)$ as well as their graded counter parts $q(2;\nu)$ and $sq(2;\nu)$ contain $sl(2)$ subalgebras generated by
\begin{equation}\label{Jyy}
J_a:=\frac{i}{8} (\gamma_a)^{\alpha \beta} M_{\alpha\beta}\ ,\quad
M_{\alpha\beta}:=q_{(\alpha}q_{\beta)}
=\tfrac{1}{2}\,( q_{\alpha}{} q_{\beta} + q_{\beta}{} q_{\alpha} )\ ,
\end{equation}
that obey
\be
[J_a,J_b] =  i\,\epsilon_{abc}\, J^c\ ,
\ee
using conventions where the matrices $(\gamma_a)^{\alpha \beta}=\epsilon^{ \beta\gamma} (\gamma_a)^{\alpha}{}_\gamma$ are normalized such that
\be \{\gamma_a , \gamma_b\}=-2\eta_{ab}\ ,\quad \rm {with}\quad \eta_{ab}={\rm diag}(-1,+1,+1)\ ,\quad \epsilon^{012}=1\ ,\ee
and the spinor indices are raised and lowered using the conventions
\be q^\alpha=\epsilon^{\alpha\beta}q_\beta\ ,\quad q_\alpha=q^\beta \epsilon_{\beta\alpha}\ ,\quad \epsilon^{\alpha\delta}\epsilon_{\beta\delta}= \delta^{\alpha}_{\beta}\ ,
\ee
together with the realization
\be ( \gamma_0 , \gamma_1 , \gamma_2 ){}^{\alpha}{}_\beta
=(-\sigma_2\, , -i \sigma_1\, , -i \sigma_3)^{\alpha}{}_\beta\ ,\quad \epsilon^{12}=\epsilon_{12}=1\ .\ee
One has
\begin{eqnarray}
  &  J^0=\frac{1}{4}\{a^+,a^-\},\qquad
  J ^\pm :=   J _1\pm  i  J _2=\frac{1}{2}(a^\pm)^2 \,,&
\label{J0,Ji}
\end{eqnarray}
where we have defined
\be q_1:=a^++ a^-\ ,\quad q_2:=i(a^+-a^-)\ ,\quad a^+=\tfrac12(q_1-i q_2)\ ,\quad a^-=\tfrac12(q_1+ i q_2)\ ,\ee
\be \label{WHalg} [a^-,a^+]=1+\nu k\ ,\quad \{k,a^\pm\}=0\ ,\quad k^2=1\ .\ee
In the $\mathbb{Z}_2$-graded case, the $sl(2)$ algebra can
be extended further to $osp(2|2)$ by taking the supercharges $Q_\a^i$ ($i=1,2$) and $so(2)$ generator $T_{12}$ to be given by \cite{Bergshoeff:1991dz}
\be Q^i_\alpha=(q_\alpha,ikq_\alpha)\ ,\quad T^{12}=-k-\nu\ ,\ee
using conventions in which $osp({\cal N}|2)$ has the following graded
commutation rules ($i=1,\dots,{\cal N}$)\footnote{The structure
coefficients of $osp({\cal N}|2)$ can be found using the realization $M_{\alpha\beta}=q_{(\alpha}q_{\beta)}$,
$Q_\alpha^i=\xi^i q_\alpha$ and $T^{ij}=i\xi^i \xi^j$ where
$q_\alpha$ obey \eqref{defy} with $\nu=0$ and $\xi^i$ are external
operators that obey $\{\xi^i,\xi^j\}=2\delta^{ij}$.}:
\be
\{Q^i_\alpha,Q^j_\beta\}=4\delta^{ij}M_{\alpha\beta}
+4\epsilon_{\alpha\beta} T^{ij}\ ,\ee
\be [M_{\alpha\beta},M^{\gamma\delta}]=8i\delta_{(\alpha}^{(\gamma}M^{\phantom{(}}_{\beta)}{}^{\delta)} \ ,\quad [T_{ij},T^{kl}]=8i\delta_{[j}^{[k} T^{\phantom{[}}_{i]}{}^{l]}\ ,\ee
\be [M_{\alpha\beta},Q_\gamma^i]=-4i\epsilon_{\gamma(\alpha}Q^i_{\beta)}\ ,\quad
[T^{ij},Q^k_\alpha]=-4i\delta^{k[i} Q^{j]}_\alpha\ .\ee
The quadratic Casimir operators
\be C_2(osp({\cal N}|2)):=J^a J_a-\frac{i}{16} Q^{\alpha i} Q_{\alpha i}+\frac{1}{32}T^{ij}T_{ij}\ .\ee
For ${\cal N}=0,1,2$, the oscillator realization gives rise to one-sided representations in various left- or right-modules, as we shall discuss below, with Casimirs
\be \label{c2sp2}C_2(sl(2))|_{\rm 1-sided}=\frac1{16}(3+2\nu k-\nu^2)\ ,\ee\be \label{c2osp12}C_2(osp(1|2))|_{\rm 1-sided}=\frac1{16}(1-\nu^2)\ ,\ee\be \label{c2osp22}C_2(osp(2|2))|_{\rm 1-sided}=0\ .\ee
The $sl(2)$ subalgebras can be extended to $sl(2)\oplus sl(2)$
by taking translations\footnote{The Lorentz generators
\begin{equation}\nonumber
L_{ab} := \epsilon_{abc} J^c\ , \qquad J_a=- \tfrac{1}{2} \epsilon_{abc} L^{bc}\ ,
\qquad [L_{ab} , L_{cd}] = i\,\eta_{bc}L_{ad} + 3 \ \rm{terms}\ ,
\end{equation}
and the translation generators
$P^a\,$ obey the commutation relations
\begin{equation}\nonumber
[J^a , J^b] =  i\, \epsilon^{abc} J_c\;, \quad
[J^a , P^b] =  i\, \epsilon^{abc} P_c\;, \quad
[P^a , P^b]= i L^{ab}\; .
\end{equation}

} $P_a$ to be realized as $P_a=J_a k$.
Instead, by tensoring with the bosonic Clifford algebra ${\rm Cliff}_1(\Gamma)$ one can take $sl(2)\oplus sl(2)\cong sl(2)\otimes {\rm Cliff}_1(\Gamma)$, with translations
\be \label{boosts} P_a= J_a \Gamma\ .\ee
We shall use the latter realization in the construction of the anyonic models, as $\Gamma$ commutes to the projectors $\Pi_\pm=\tfrac12(1\pm k)$ used to define the tensorial, fractional-spin and Lorentz-singlet representations making up the fractional-spin gravity model.

\subsection{Representation of $Aq(2;\nu)$ in Fock space: $\check{Aq}(2;\nu;\tau)$}
\label{subsec:RepresentationAq}

Following \cite{Plyushchay:1997mx}, one can represent the Wigner-deformed Heisenberg  oscillator
algebra \eqref{WHalg} in terms of undeformed oscillators obeying
\begin{equation}
    [ b^-, b^+]=1\,.
\end{equation}
To this end, one represents the elements $f$ of the oscillator algebras by operators $\check f$ acting in a Fock space ${\cal F}$,
\be \mathcal{F}=\bigoplus_{n=0}^\infty \mathbb C\otimes |n\rangle\ ,\quad |n\rangle=\frac1{\sqrt{n!}} (\check b^+)^n|0\rangle\ ,\quad \check b^-|0\rangle=0\ ,\ee
\be (\check N-n)|n\rangle=0\ ,\quad \check N:=\check b^+ \check b^-\ .\ee
In the Fock space, the deformed oscillators and the Klein operator can be represented by the following non-linear constructs:
\begin{eqnarray}
    &&\check a^+= \left(\check G \;\sqrt{1+\frac{\nu}{\check N}}\,
    \check {\Pi}_- + \check H
   \,\check {\Pi}_+\right)\check b^+\,,\\[5pt]
   &&\check a^- =
     \check b^- \left(\check G^{-1}\sqrt{1+\frac{\nu}{\check N}}\, \check \Pi_-+  \check H^{-1}\,
    \check \Pi_+\right)\,,\\[5pt]&& \check k=(-1)^{\check N}\ ,\quad \check G=G(\check N)\ ,\quad \check H=H(\check N)\ ,
\end{eqnarray}
where $\check \Pi_\pm=\frac{1}{2}(1\pm \check k)$ such that
$\check J^0\equiv \tfrac14 \{\check a^+,\check a^-\}=\tfrac12 \check N+\tfrac14(1+\nu)$ as in \eqref{J0} with $\varepsilon=+1$.
In particular, taking $ \check{H} = 1 $
and $\check{G} = ( 1 +\frac{\nu}{\check{N}})^{\tau}$
where $\tau\in\Real$ one has
\begin{equation}\label{a+a-F1}
    \check a^+ =\left(1+\frac{\nu}{\check N}\right)^{1/2+\tau}
    \;\,\check b^+\check {\Pi}_+ +
    \check b^+\check {\Pi}_-\,,\end{equation}\begin{equation} \label{a+a-F2} \check a^- =
    \left(1+\frac{\nu}{\check N+1}\right)^{1/2-\tau}
    \;\, \check b^- \check \Pi_-+ \check b^-
    \check \Pi_+\,,
\end{equation}
with formal inverse
\begin{equation}\label{b+}
    \check b^+ =\left(1+\frac{\nu}{\check N}\right)^{-1/2-\tau}
    \;\,\check a^+\check {\Pi}_+ +
    \check a^+\check {\Pi}_-\,,\end{equation}\begin{equation} \label{b-} \check b^- =
    \left(1+\frac{\nu}{\check N+1}\right)^{-1/2+\tau}
    \;\,\check a^- \check \Pi_-+\check a^-
    \check \Pi_+\,.
\end{equation}
We denote the resulting representation of $Aq(2;\nu)$ in ${\cal F}$ by $\check{Aq}(2;\nu;\tau)$, which is thus the associative algebra consisting of arbitrary polynomials in $\check{a}^\pm$ and $\check{k}$ as given above with parameter $\tau\in \Real$.

For $\nu=0$ (and all $\tau$) one has $\check a^\pm=\check b^\pm$ and the representation of $Aq(2;0)$ in ${\cal F}$ is unitary if one chooses the hermitian conjugation rule
$(\check b^+)^{\check{\dagger}}= \check b^{-}$\,.
One has the standard sesquilinear form defined by
\be (|0\rangle)^{\check\dagger}:=\langle 0|\ ,\quad \langle 0|0\rangle:=1\ .\ee
Thus $(|n\rangle)^{\check\dagger} = \langle n|$
and the Klein operator is realized in the Fock space for all $\nu$  by
\be
\label{klein}
\check k= \sum_{n\geqslant 0} (-1)^n  | n \rangle \langle n| \,.
\ee
For finite $\nu$ there exist hermitian conjugation operations of the form \cite{Plyushchay:1997mx}
\be
\label{conjn}
(\check a^\pm)^{\check \dagger} =\check C^{-1}\,\check a^\mp \check C,\qquad
(\check k)^{\check \dagger}=\check C^{-1}\,\check k\, \check C\ ,
\ee
such that
\begin{equation}\label{Yconj}
(\check q_\alpha)^{\check \dagger} = \check C^{-1} \,\check q_\alpha\, \check C \,,
 \end{equation}
where the conjugation matrix $\check C\in {\rm End}({\cal F})$ depends on $\nu$, or rather, as we shall see, the integral part $[\nu]\,$.
We may further require that
\be \label{Creal}(\check f^{\check\dagger})^{\check \dagger}=\check f\quad \mbox{for any $\check f\in{\rm End}({\cal F})$}\quad \Leftrightarrow\quad  \check C^{\check \dagger} = \check C\ .
\end{equation}
Imposing also
\be \check C|0\rangle=|0\rangle\,,\ee
it follows that the sesquilinear form
\be \langle  \xi| \check C |\chi \rangle\equiv (|\xi \rangle)^{\check\dagger} \check C |\chi\rangle \,,
\end{equation}
is invariant under similarity transformations generated by the elements $\check f\in {\rm End}({\cal F})$ that satisfy the reality condition
\be  \check f^{\check \dagger} \,=-\check C^{-1}\check f \check C \,,
\end{equation}
\emph{viz.}
\begin{equation}
\langle\widetilde{\xi}|\check C|\widetilde{\chi}\rangle = \langle{\xi}|\check C|{\chi}\rangle
,\quad \hbox{where} \quad
|\widetilde{\xi}\rangle=\exp(\check f)|\xi\rangle\,, \quad
|\widetilde{\chi}\rangle=\exp(\check f)|\chi\rangle\,.
\end{equation}
One may further restrict
\be \label{Cunimodular} \check C^2=1\quad \Leftrightarrow\quad \tau=0\ ,\ee
for which one has
\begin{equation}\label{a+a-}
\quad  \check a^+= \sum_{n\geq 0} \sqrt{[n+1]_\nu}\; |n+1\rangle \langle n | ,
 \qquad \check a^-=
\sum_{n\geq 0} \sqrt{[n+1]_\nu}\; |n\rangle \langle n+1| ,
\end{equation}
where
\begin{equation}
 [n]_\nu :=n+\frac{1}{2}(1-(-1)^n)\nu \, .
\end{equation}
One may choose a diagonal conjugation matrix
\be \label{C1} \check C =\sum_{n\geq 0} C_n |  n \rangle \langle n |  \, .
\end{equation}
Factoring out the relations $\check a^-|0\rangle= 0 = (\check k-1)|0\rangle\,$ yields a generalized Verma module spanned by
\be |n):=(\check a^+)^n|0\rangle\,, \quad n=0,1,2,...\ ,
\ee
which are non-normalized eigenstates of $\check N\,$.
The emergence of singular vectors, that is, states $|n)$ with $n>0$ that are annihilated by $\check a^-$, is associated with the existence of finite-dimensional representations of the
Wigner-deformed Heisenberg algebra, that is, realizations of the algebra
in terms of finite-dimensional matrices.
Defining
\begin{equation}\label{bras}
 (n|:=\langle 0| (\check a^-)^n=\langle 0| ((\check a^+)^n)^{\check \dagger} \check C
 \,,\quad n=0,1,2,...\ ,
 \end{equation}
such that $(n'|n)\equiv \langle 0| (\check a^-)^{n'} \check C (\check a^+)^n|0\rangle$,
it follows that if $|n)$ is a singular vector then $(n'|n)=0$ for all $n'$.
As
\begin{eqnarray}\label{sprod}
(n'|n) =\delta_{n',n}\:[n]_\nu!  \ , \qquad
[n]_\nu!  \;  := \;  \: \prod_{m=1}^n [m]_\nu\: , \qquad n',n \geqslant 0\,,
\end{eqnarray}
the following cases arise:
\begin{enumerate}
\item[I.] \underline{$ \nu>-1$}: In this \emph{unitary} regime,
the matrix elements $(n|n)=[n]_\nu ! >0 $ for all $n\,$, and hence
\be \check C=1\ ,\qquad \check a^\pm=(\check a^\mp)^\dagger\ .\ee
The representation of the deformed oscillators in $\mathcal{F}$ is thus unitary \cite{Plyushchay:1997mx,Plyushchay:1997ty}.
\item[II.] \underline{$\nu=-1$}: In this \textit{hyper-critical} case, which is also unitary, one has
\be
\check C=1\ ,\qquad \check a^+|0\rangle =0\ ,\quad \quad \check a^-|1\rangle=0\ ,
\ee
and the representation $\check{Aq}(2;-1;0)$ decomposes into
\be
\check{Aq}(2;-1;0)= gl(1)\oplus \check{Aq}(2;1;0)\ ,
\ee
that is, $\mathcal F$ decomposes under $Aq(2;-1)$ represented as $\check{Aq}(2;-1;0)$ into a
singlet $|0\rangle$ and an infinite-dimensional unitary representation of $Aq(2;1)$ in $\bigoplus_{n\geqslant 1} \Comp |n\rangle$ --- as shown below Eq. \eqref{flip}.
\item[III.] \underline{$\nu=-2\ell-1,\:\: \ell=1,2,...$}: In these \emph{critical} cases, one has
 \begin{equation}\label{lhw}
 \check a^+ | 2\ell \rangle =0  \, ,\quad  \check a^- |2\ell+1 \rangle =0 \,,
\end{equation}
and
\begin{eqnarray}
&& {\rm sign}[(n|n)]=\left\{ \begin{array}{ll}
                        \cos\Big(\frac{n \pi}{2} \Big) -
\sin\Big(\frac{n \pi}{2} \Big)  & 0\:\leqslant \:n\: \leqslant 2 \ell \,, \\[5pt]
0  & n\: \geqslant 2 \ell+1  \,,
                 \end{array}
\right.
\end{eqnarray}
where $sign(x):=x/|x|$ for $x\neq0$ and $sign(0):=0\,$.
It follows that
\begin{eqnarray}
\label{Cnuni}
\check C =\sum_{n} C_{n} \, |n\rangle \langle n|  \,, \qquad C_{n} =\left\{ \begin{array}{ll}
                        \: {\rm sign}[(n|n)]  & 0\:\leqslant \:n\: \leqslant 2 \ell
\\[5pt]
1  & n\: \geqslant 2 \ell+1  \,,
                 \end{array}
\right.
\end{eqnarray}
and that ${\cal F}$ decomposes into two irreducible representations of the deformed oscillators,
\be
\mathcal{F}=\mathcal{F}_{\rm f}\, \oplus \mathcal{F}_\infty\ ,\qquad \mathcal{F}_{\rm f}=\bigoplus_{n=0}^{2\ell} ~\Comp\otimes |n\rangle \ ,\qquad \mathcal{F}_{\infty}=\bigoplus_{n\geqslant 2\ell+1}\Comp\otimes |n\rangle\ .
\ee
Indeed, the projectors
\begin{equation}\label{nullproj}
 \mathcal{P}_{\rm f}:=\sum_{n=0}^{2\ell} |n\rangle \langle n|  \,, \qquad
\mathcal{P}_{\infty}:=\sum_{n=2\ell +1}^{\infty} |n\rangle \langle n|\, ,
\end{equation}
commute with $(\check a^\pm,\check k)$ iff $\nu$ is critical or hyper-critical, and hence
\be \check a^\pm=\check a^\pm_{\rm f}+\check a^\pm_\infty \ ,\quad \check k=\check k_{\rm f}+\check k_{\infty}\ ,\quad \check C=\check C_{\rm f}+\check C_\infty\ ,\ee
\begin{equation}
\check a^\pm_{ \stackrel{{\rm f}} { \infty }} := \mathcal{P}_{ \stackrel{{\rm f}} { \infty }}\, \check a^\pm \, \mathcal{P}_{ \stackrel{{\rm f}} { \infty }}\, , \qquad
\check k_{ \stackrel{{\rm f}} { \infty }} := \mathcal{P}_{ \stackrel{{\rm f}} { \infty }}\, \check k \, \mathcal{P}_{ \stackrel{{\rm f}} { \infty }}\, ,\qquad \check C_{ \stackrel{{\rm f}} { \infty }} := \mathcal{P}_{ \stackrel{{\rm f}} { \infty }}\, \check C \, \mathcal{P}_{ \stackrel{{\rm f}} { \infty }}\, ,
\end{equation}
obey
\begin{equation}\label{dhafinf}
\big[\check a^-_{ \stackrel{{\rm f}} { \infty }},\check a^+_{ \stackrel{ {\rm f} } { \infty }} \big]=1-(2\ell+1) \check k_{ \stackrel{{\rm f}} { \infty }} \, , \qquad
\{\check k^{\phantom{\pm}}_{ \stackrel{{\rm f}} { \infty }},\check a^\pm_{ \stackrel{{\rm f}} { \infty }}\big \}=
0 \,,
\end{equation}
and the hermicity conditions
\begin{equation}
\check a^\mp_{\rm f}=\check C^{\phantom{\pm}}_{\rm f}\,
(\check a^\pm_{\rm f})^{\check \dagger} \, \check C^{\phantom{\pm}}_{\rm f}\,,\qquad
\check k^{\phantom{\pm}}_{\rm f}=\check C^{\phantom{\pm}}_{\rm f}\, \check k_{\rm f}^{\check \dagger} \, \check C^{\phantom{\pm}}_{\rm f}=\check k_{\rm f}^{\check \dagger} \,,\qquad \check a^\mp_{\infty}= \,
(\check a^\pm_{\infty })^{\check \dagger} \,, \qquad \check k^{\phantom{\pm}}_{\infty}=\check k_{\infty}^{\check \dagger} \,.
\end{equation}
In terms of the bra-ket basis, one has
\begin{equation}\label{a+a-f}
    \begin{array}{c}
    \check a^+_{{\rm f}}= \sum_{n= 0}^{2\ell} \sqrt{[n+1]_\nu} |n+1\rangle \langle n | ,\qquad \check a^-_{\rm f}=
\sum_{n= 0}^{2\ell} \sqrt{[n+1]_\nu} |n\rangle \langle n+1| \, ,\\[10pt]
\check k_{\rm f}= \sum_{n= 0}^{2\ell}\,  (-1)^n \,  |
n
\rangle \langle n| \,,\qquad \check C_{\rm f} =\sum_{n=0}^{2\ell} \left( \cos\Big(\frac{n \pi}{2} \Big) -
\sin\Big(\frac{n \pi}{2} \Big) \right) \: |n\rangle \langle n|  \,,

\end{array}
\end{equation}
and
\begin{equation}\label{a+a-inf}
 \begin{array}{c}
    \check a^+_{\infty}= \sum_{n \geqslant 2\ell+1} \sqrt{[n+1]_\nu} |n+1\rangle \langle n | ,\qquad \check a^-_{\infty}=
\sum_{n \geqslant 2\ell+1} \sqrt{[n+1]_\nu} |n\rangle \langle n+1| ,\\[10pt]
\check k_{\infty}= \sum_{n \geqslant 2\ell+1}  \, (-1)^n \, |
n
\rangle \langle n| \,,\qquad \check C_\infty=1\ .
\end{array}
\end{equation}
Thus, $(\check a^\pm_{\rm f},\check k_{\rm f})$ provide a finite-dimensional non-unitary representation of the Wigner--Heisenberg algebra with deformation parameter $\nu = -2\ell-1$ whose enveloping algebra is isomorphic to $gl(2\ell+1)$, while $(\check a^\pm_{\infty},\check k_{\infty})$ provide an infinite-dimensional unitary representation of the Wigner--Heisenberg algebra with deformation parameter $2\ell+1$, as can be seen from
\be  \check k_{\infty}|2\ell+1\rangle=-|2\ell+1\rangle\ ,\label{flip}\ee
which implies that the redefinition $\check k_{\infty} \rightarrow - \check k_{\infty} $ yields a representation of $Aq(2;2\ell+1)$ on ${\cal F}_\infty$.
Thus, at critical $\nu$ one has
\be \check{Aq}(2;-2\ell-1;0)\cong gl(2\ell+1)\oplus \check{Aq}(2;2\ell+1;0)\ .\label{caseIII}\ee

\item[IV.] \underline{$\nu \, < \, -1\,$, $\nu\notin \{-3,-5,\dots\}$}: For these non-critical values, the representation of the deformed oscillators in ${\cal F}$ is irreducible and \emph{non-unitary}, as can be seen from
\begin{eqnarray}
&& sign[(n|n)]=\left\{ \begin{array}{ll}
                        \cos\Big(\frac{n \pi}{2} \Big) -
\sin\Big(\frac{n \pi}{2} \Big)  & 0\:\leqslant \:n\: \leqslant 2 \ell \\[5pt]
(-1)^{\ell+1}  & n\: \geqslant 2 \ell+1  \,.
                 \end{array}
\right.
\end{eqnarray}
where $\ell$ is the positive integer defined by that $2\ell+1$ is the supremum of odd integers less than $|\nu|$.
The conjugation matrix is thus given by

\begin{equation}\label{CII}
\check C =\sum_{n=0} C_{n} \, |n\rangle \langle n|  \,, \qquad C_{n}= \: {\rm sign}[(n|n)] \,.
\end{equation}
\end{enumerate}

\subsection{Fractional-spin representations of Lorentz and AdS algebras}

The representation of the Lorentz algebra \eqref{standardso12} in terms
of the deformed oscillators is reducible and it can be projected as in
\eqref{anyonso12}. On top of this reducible structure there may arise another
one depending on the value of the deformation parameter $\nu\,$. This will
affect the field content in the higher-spin Chern--Simons theory
presented in Section \ref{sec:CS}.

{}From \eqref{a+a-} it follows that the representation of the Lorentz generators \eqref{J0,Ji} in the Fock
space is given by,
\begin{eqnarray}
 \check J^0 &=& \sum_{n\geqslant 0} \Big(\frac{n}{2}+\frac{1+ \nu}{4} \Big) | n
\rangle \langle n|\,,\label{J0F}\\[5pt]
 \check  J^- &=& \sum_{n\geqslant 0} \sqrt{[n+2]_\nu[n+1]_\nu}  | n\rangle \langle  n+2|\,,\\[5pt]
 \check  J^+ &=& \sum_{n\geqslant 0} \sqrt{[n+2]_\nu[n+1]_\nu}  |
n+2
\rangle \langle n|\,.\quad
\end{eqnarray}
The quadratic Casimir operator \eqref{c2sp2} factorizes into
\begin{equation}\label{JJ}
   C_2(sp(2)|{\cal F})\equiv -\check \alpha (\check \alpha -1)\,
\quad\hbox{with}\quad
    \check \alpha =\frac{1}{4}(2+\nu-\check k) \,.
\end{equation}
Since the Klein operator take place in this expression, the value of the
Casimir operator does not take a fixed value \cite{Vasiliev:1989re}.
The Fock space thus decomposes into two invariant eigenspaces of $\check k$,
\begin{equation} \label{calF+-}
\mathcal{F}_\pm=\check\Pi_\pm \mathcal{F}=\bigoplus_{n=0}^\infty \mathbb C\otimes
|2n+\tfrac{1}{2}(1\mp1)\rangle\,,
\end{equation}
where $\check \Pi_\pm$ are the projectors defined in \eqref{projector}.
The projected Lorentz generators and spins are given by
\begin{equation}\label{Jirrep1}
  \check J ^\pm_a := \check \Pi_\pm   \check J _a\,, \quad \eta^{ab}  \check J ^\pm_a \check J^\pm_b =-j^\pm(j^\pm-1)\, \check \Pi_\pm\,,
\quad j^+=\frac{1}{4}(1+\nu)\,,
\quad j^-=\frac{1}{4}(3+\nu)\,.
\end{equation}
The spins of the odd and even representations differ by half a unit,
\be j^--j^+=\frac12\ ,\ee
 thus forming superpartners.
Hence, in the non-critical case, the Fock space carries two irreducible representations
 of the Lorentz algebra.
In the critical cases, there is a further sub-decomposition into a finite-dimensional and an infinite-dimensional irrep due to the additional projectors \eqref{nullproj}, \emph{viz.}
\begin{equation}\label{Jirrep2}
  \check J ^{(f)\pm}_a := \mathcal{P}_{\rm f} \check \Pi_\pm   \check J _a
\,, \qquad   \check J ^{(\infty)\pm}_a := \mathcal{P}_{\infty} \check \Pi_\pm \check J_a\,,
\end{equation}
with the spin in each irreducible sector given in the Table \ref{Tirrep}.
This additional reducibility is reflected in the symmetry of the Lorentz Casimir
operator under $j\rightarrow
1- j$, yielding different representations of the Lorentz algebra for which
\be j^+-j^-=\frac12\ .\ee
Thus, for $\nu=-3,-5,-7,...$ one has two finite non-unitary and two infinite
dimensional unitary representations of the Lorentz algebra.
In the hyper-critical case $\nu=-1$, the finite dimensional subspace contains only one state, the ground state, which is invariant under the action of the full $Aq(2;-1)$ algebra represented as $\check{Aq}(2;-1;0)$.
Indeed, the representation $\check{Aq}(2;-1;0)$ of the algebra $Aq(2;-1)$ is unitary since the conjugation matrix is the identity.

\begin{table}[ht]
\begin{center}
\begin{tabular}{|c|c|c|}
 \hline
 $\nu$ & Irreducible subspaces & Lorentz spin 
\\\hline\hline
&&\\[-10pt]
non-critical \\
$\nu > -1$ &
$\begin{array}{c}
|2n \rangle ,\quad n=0,1,2,...\\[4pt]
|2n+1\rangle ,\quad n=0,1,2,...
\end{array}$
&
$\begin{array}{c}
j^+=(1+\nu)/4 \\[4pt]
j^-=(3+\nu)/4
\end{array}$\\
&&\\[-10pt]
\hline
&&\\[-10pt]
\begin{tabular}{l}
 critical \\
$\nu=-(2\ell+1),$\\
$\ell=1,2,...$
\end{tabular}
&
\begin{tabular}{l}
${\cal F}_{\rm f}$ $\left\{ \begin{array}{c}
|2n\rangle ,\quad n=0,1,2,...,\ell\qquad \quad \: \\[4pt]
|2n+1\rangle ,\quad n=0,1,2,...,\ell-1
\end{array} \right.$\\[4pt]
${\cal F}_{\infty}$ $\left\{ \begin{array}{c}
|2n\rangle ,\quad
n=\ell+1,\ell+2,...\\[4pt]
|2n+1\rangle,\quad n=\ell,\ell+1,...
\end{array} \right.$
\end{tabular}
&
$\begin{array}{c}
j^{(f)+}=-\ell/2 \\[4pt]
j^{(f)-}=1/2-\ell/2 \\[4pt]
j^{(\infty)+}=1+\ell/2 \\[4pt]
j^{(\infty)-}=1/2+\ell/2
\end{array}$\\
&&\\[-10pt]
\hline
&&\\[-10pt]
\begin{tabular}{l}
 hyper critical \\
$\nu=-1$
\end{tabular}
&
\begin{tabular}{l}
${\cal F}_{\rm f}$ $\left\{ |0\rangle \right. ,$ \\[4pt]
${\cal F}_{\infty}$ $\left\{ \begin{array}{c}
|2n\rangle ,\quad
n=1,2,...\\[4pt]
|2n+1\rangle,\quad n=0,1,...
\end{array} \right.$
\end{tabular}
&
$\begin{array}{c}
j^{(f)+}=0 \\[4pt]
j^{(\infty)+}=1 \\[4pt]
j^{(\infty)-}=1/2
\end{array}$
\\
&&\\
\hline
\end{tabular}
\end{center}
\caption{Representing $sl(2)$ in terms of Wigner-deformed Heisenberg oscillators in a standard Fock
space yields reducible representations of $sl(2)$.
The first column contains the values of the deformation parameter $\nu$ in the
Wigner-deformed Heisenberg algebra.
The second column contains the corresponding $sl(2)$-irreducible subspaces of the Fock space.
The third column contains the corresponding values of the spins $j$, \emph{i.e.} the $J^0$ eigenvalue
of the lowest weight state in each $sl(2)$-irrep.
}
\label{Tirrep}
\end{table}

\vspace{5mm}

The classification of the unitary irreducible
representations of $SL(2,\mathbb{R})$ was first done by Bargmann
\cite{Bargmann:1946me}.
Comparing with the unitary irreducible representations of $sl(2)$
in Fock space by Barut and Fronsdal \cite{Barut:1965}
and adapting the notation to this paper, we see that the unitary irreducible representations appearing 
in the non-critical case $\nu > -1$  above furnish the discrete series ${\cal D}^+(j^\pm)\,$. For  $\nu <-1$, but non-critical, these representations are still of a discrete type, but non-unitary. In reference \cite{oai:arXiv.org:1001.0274} it was shown that the latter were essential for the construction of anyon wave equations possessing standard boson/fermion limits.

\subsection{Irreducible representations of $sl(2)$ in two-sided Fock space}

A generic operator $\check f\in \check{Aq}(2;\nu;\tau)$ can thus be represented in ket-bra form as
\begin{equation} \label{op}
\check f= \sum_{n,m\geq 0}  f^{mn} | m \rangle \langle
n| \,,
\end{equation}
where the matrix $\{f^{mn}\}$ becomes block diagonal for critical $\nu$ if $\tau=0$.
The operators $| m \rangle \langle n| $ are products of harmonic oscillator states with $\nu$-dependent spin,
\be \check J^0 |m\rangle = s_m |m\rangle\, ,\qquad  \langle m | \check J^0  = \langle m | s_m \, , \qquad s_m = \tfrac{m}2+\tfrac{1+ \nu}{4} \, ,\ee
transforming under a $2 \pi$ rotation by an anyonic statistical phase,
\begin{equation}
\exp(i 2\pi    \check J^0 ) | m\rangle   = e^{i\pi(m+\tfrac{1+\nu}{2})}  \; | m\rangle \,.
\end{equation}
The tensor product $| m \rangle \langle n |$, which transforms
in the adjoint representation of the rotation group generated by $\check J^0$,
\begin{equation}\label{rotketbra}
 [\,   \check J^0, | m
\rangle
\langle n|\,]= (s_m-s_n)\, | m \rangle \langle n|  \,, 
\end{equation}
transforms by a standard phase  $+ 1$ or $-1$, \emph{viz.}
\be
\exp(i2\pi \,   \check J^0)| m
\rangle
\langle n| \exp(-i2\pi \,   \check J^0)  =(-1)^{m-n} | m
\rangle \langle n| \,,
\ee
hence corresponding to bosonic or fermionic statistics.

For the alternative choice of the Lorentz generators \eqref{Jirrep1}, the  ket-bra products transform as follows:
 \begin{equation}
 [\,   \check J ^{\pm}_0, | m
\rangle
\langle n|\,]= \tilde{s}^\pm_{m,n} \, | m \rangle \langle n| \,, \qquad \tilde{s}^\pm_{m,n}:= s_m \frac{(1\pm(-1)^m)}2 - s_n \frac{(1\pm(-1)^n)}2   \,.
\end{equation}
It follows that if $n$ and $m$ have different parity then $| m \rangle \langle n |$ will transform under either the left or right action of the rotation group, and hence their spin will have a $\nu$-dependent fractional component.
This observation indicates that, in order to include particles with fractional spin into the higher-spin connection, one needs to identify the Lorentz connection, which activates the local rotation symmetry, with the generators  $J ^{\pm}_a$.
The $AdS$ algebra $so(2,2)\cong sl(2)\oplus sl(2) $ is obtained by  doubling the algebra as in \eqref{boosts}.

\subsection{Polynomial versus Fock-space bases}\label{Sec:bases}

In order to construct the fractional-spin algebras, we start from the anyon representations\footnote{To our best understanding, the complete classification of all possible representations of $Aq(2;\nu)$ is an open problem. Indeed, the classification of  infinite-dimensional irreducible representations of finite-dimensional Lie algebras is an active field in pure mathematics \cite{KnappOverview}.
Two key differences between finite- and infinite-dimensional irreps is that the former are completely decomposable and can be labelled by the Casimir operators, while the latter, which can exhibit different branches of indecomposable structures, cannot be labelled faithfully only by Casimir operators.
Additional ``Langlands parameters'' \cite{KnappOverview} are thus required to distinguish the infinite-dimensional irreducible representations, such as the parameter $\tau$ introduced in Eqs. \eqref{a+a-F1} and \eqref{a+a-F2}.}
$\check{Aq}(2;\nu;\tau)$ of $Aq(2;\nu)$ in the Fock space ${\cal F}$, which have spins $\tfrac14(1+\nu)$ and $\tfrac34(1+\nu)$.
As discussed in subsection \ref{Aqsec}, the actions of $Aq(2;\nu)$ on itself from the left or from the right provide faithful representations of the algebra.
Moreover, as we have seen in subsection \ref{subsec:RepresentationAq}, the representation $\check{Aq}(2;\nu;\tau)$ of $Aq(2;\nu)$ on ${\cal F}$ is isomorphic to ${\rm End}({\cal F})$ for generic values of $\nu$; for critical values, the algebra $\check{Aq}(2;\nu;\tau)$ becomes a subalgebra of ${\rm End}({\cal F})$ with an (in)decomposable structure determined by $\tau$.
The algebras $\check{Aq}(2;\nu;\tau)$ are isomorphic to subalgebras $\rhoF(\check{Aq}(2;\nu;\tau))$ inside the non-polynomial completion \eqref{fqk} $\overline{Aq}(2;\nu)$ of $Aq(2;\nu)$ by means of the deformed-oscillator
realization of the vacuum-to-vacuum projector, in accordance with \eqref{rhoF}.

The Fock space ${\cal F}$, viewed as an $sl(2)$ module, decomposes into two fractional-spin representations in the discrete series
\cite{Bargmann:1946me}.
In these representations the spin operator $J^0$ acts diagonally with real-valued eigenvalues.
The Fock-space module can thus be identified with $\rhoF(\check{Aq}(2;\nu;\tau))$ viewed as either a left module or a right module.
On the other hand, the separate left and right actions in $Aq(2;\nu)$ also give rise to $sl(2)$ modules but of a different type since the only generators of $Aq(2;\nu)$ that acts diagonally on itself from one side is the identity $1$ and the Kleinian $k$.

To illustrate the inequivalence between $Aq(2;\nu)$ and $\rhoF(\check{Aq}(2;\nu;0))$ for critical values
$\nu=-2\ell-1$, $\ell=0,1,2,\dots$, one may consider the $++$-projection defined in \eqref{projector}.
For this projection, one has
\be
\label{aq1}\check{Aq}(2;-2\ell-1;0)_{++}\cong gl(\ell+1)\oplus \check{Aq}(2;2\ell+1;0)_{++}\ ,
\ee
in agreement with the result obtained in \eqref{caseIII} (as a consequence of the existence of new projector operators \eqref{nullproj} which split the Fock space into two sectors of finite and infinite dimension).
On the other hand, the action of $Aq(2;-2\ell-1)_{++}$ on itself exhibits an indecomposable structure of the form
\cite{Vasiliev:1989re}
\be
\label{aq2}
Aq(2;-2\ell-1)_{++} = \frac{Aq(2;-2\ell-1)_{++}}
{Aq'(2;-2\ell-1)_{++}} \supset\!\!\!\!\!\!+ Aq'(2;-2\ell-1)_{++}\ ,
\ee
where the ideal $Aq'(2;-2\ell-1)_{++}$ is spanned by $\Pi_+ q_{(\a_1}\cdots q_{\a_{2n})}$ with $n=\ell+1,\ell+2,\dots$ and the quotient
\be
\frac{Aq(2;-2\ell-1)_{++}}
{Aq'(2;-2\ell-1)_{++}}\cong gl(\ell+1)
\ee
is spanned by $\Pi_+ q_{(\a_1}\cdots q_{\a_{2n})}$ with $n=0,1,\dots,\ell$
(modulo elements in $Aq'(2;\nu)_{++}\,$).
Thus, the indecomposable structures of $Aq(2;-2\ell-1)_{++}$ and
$\check{Aq}(2;\nu;-\frac12)_{++}$ are of different types,
with $Aq(2;-2\ell-1)_{++}$ containing a non-trivial ideal and
$\check{Aq}(2;-2\ell-1;0)_{++}$ having a block-diagonal structure.
By choosing other values for $\tau$ it is possible to alter the indecomposable structure
of $\check{Aq}(2;\nu;\tau)_{++}$ in critical limits.
In particular, for $\tau=-\frac12$ it follows that $\check{Aq}(2;-2\ell-1;-\frac12)_{++}$
has an indecomposable structure of the same type as $Aq(2;-2\ell-1)_{++}$,
in the sense that both algebras have infinite-dimensional ideals and coset algebras given by
$gl(\ell+1)\,$.
Note, however, that $\check{Aq}(2;-2\ell-1;-\frac12)_{++}$ and $Aq_{++}(2;-2\ell-1)_{++}$ are not isomorphic as $sl(2)$ representations since the spin operator is diagonal in the former space but not in the latter.

In fact, the above conclusions do not change considerably if one removes the $++$ projection.
In a generalization of Feigin's notation \cite{Feigin88}, we define
\be gl(\lambda;J;\tau):=\left.\left.\frac{{\rm Env}(sl(2))}{I(\l)}\right\downarrow_{J}\right|_{\tau}\ ,\ee
where $I(\l)$ is the ideal generated by $C_2(sl(2)) + \lambda(\lambda-1)$;  $(\cdot)\downarrow_J$ indicates that the elements in $(\cdot)$ are given
in a basis where the generator $J\in sl(2)$ acts diagonally from both sides;
and $\tau$ parameterizes the indecomposable structure.
In particular, Feigin's original construction was performed in the basis of monomials in the generators of $sl(2)$ in which no generator $J$ can be diagonal;
we denote this particular basis by $gl(\lambda;-;-)$.
With this notation, it follows that
\be
Aq(2;\nu)_{\sigma\sigma}\cong gl(\tfrac{2+\nu-\sigma}{4}\,;-;-)\ ,
\ee
\be
\check{Aq}(2;\nu;\tau)_{\sigma\sigma}\cong  gl(\tfrac{2+\nu-\sigma}{4}\, ; \check J^0;\tau)\ ,
\ee
which are thus infinite-dimensional algebras for generic $\nu$ with critical limits given by semi-direct sums of a finite-dimensional and an infinite-dimensional sub-algebra with ideal structure controlled by $\tau$.
Using this notation, one can write Eqs. \eqref{aq1} and \eqref{aq2} as
\be
\label{gl2}
Aq(2;-2\ell-1)_{++}\cong gl(\ell+1;-;-\tfrac{1}{2})\cong
gl(\ell+1)\supset\!\!\!\!\!\!+ gl(-\ell;-;-\tfrac{1}{2})\ ,
\ee
\be \label{gl1}\check{Aq}(2;-2\ell-1;0)_{++}\cong gl(\ell+1;J_0;0)\cong gl(\ell+1)\oplus gl(-\ell;J_0;0)
\ .\ee

\subsection{Real forms of $Aq(2;\nu)$ and related Lie (super) algebras}

There are two ways to impose reality conditions on the elements of the derived Lie algebras of $Aq(2;\nu)$ using either hermitian conjugations or complex conjugations, also known as star-maps, giving rise to infinite-dimensional analogs of the real forms $gl(n;\Real)$ and $u(p,q)$ of $gl(n;\mathbb C)$, respectively.
Various such conjugations can be obtained by combining inner automorphisms $\varphi={\rm Ad}_S$ of $Aq(2;\nu)$ with the basic hermitian conjugation operation $\dagger$ defined in \eqref{conjn} and the linear anti-automorphism $\tau$ defined by
\begin{equation}\label{tau1}
  \tau(f_1{} f_2):=\tau(f_2){}\tau(f_1)\ ,\quad \tau( \beta q_\a):=i\beta q_
  a\ ,\quad \tau(\beta|0\rangle):=\beta\langle 0|\ ,
\end{equation}
where $\beta\in\mathbb{C}$ (and we note that $\tau( a^\pm)=a^\mp$).
As for the associative algebra itself, its real forms require star-maps;
the real form
\be Aq_{\varphi}(2;\nu;\Real):=\left\{ f\in Aq(2;\nu): \varphi f^\ast =f\right\}\ ,\qquad f^\ast:=\tau(f^\dagger)\ .\ee
Assuming that $((f^\ast)^\ast)=f$ for all $f$ it follows that $S S^\ast=1$.
Assuming furthermore that $S=\tilde S^2$ and that $\widetilde S \widetilde S^\ast=1$ it follows that if $f^\ast=\varphi(f)$ then $({\rm Ad}_{\widetilde S}(f))^\ast={\rm Ad}_{\widetilde S}(f)$, that is,
\be Aq_{\varphi}(2;\nu;\Real)\cong Aq_{\rm Id}(2;\nu;\Real):=\left\{ f\in Aq(2;\nu): f^\ast=f\right\}\ .\ee
Starting from $Aq(2;\nu;\Real)$ various real forms of $slq(2;\nu;\Real)$ and $sq(2;\nu;\Real)$ can then be reached by generalizations of the Weyl unitarity trick as follows:
\begin{eqnarray} lq(2;\nu;\Real)&&:= \left\{ h\in lq(2;\nu)\right\}\cap Aq_{\rm Id}(2;\nu;\Real)\ ,\\[5pt] uq(2;\nu)&&:= \left\{ h=f+ig\,,\  f,g\in lq(2;\nu)|\  \tau(f) =-f\,,\ \tau(g)=g\right\}\cap Aq_{\rm Id}(2;\nu;\Real)\qquad\\&&=\{h\in lq(2;\nu)|\ h^\dagger=-h\}\ ,\\[5pt]
hosl(2|2;\nu)&&:= \left\{ h\in q(2;\nu)\right\}\cap Aq_{\rm Id}(2;\nu;\Real)\ ,\\[5pt]
 hosp(2|2;\nu)&& := \left\{h\in q(2;\nu)|h^\dagger = - i^{{\rm deg}(h)} h\,,\right\}\ .\end{eqnarray}
In the two cases projected out by hermitian conjugation, their Fock space representations take the form
\be
\check uq(2;\nu;\tau):=\left\{\check h\in \check lq(2;\nu;\tau): \check h^{\check \dagger}
=-\check C \check h \check C\right\} \, ,
\ee
\be
\check hosp(2|2;\nu;\tau):=\left\{ \check h\in \check q(2;\nu;\tau): \check h^{\check \dagger}
=-i^{{\rm deg}(\check h)} \check C \check h \check C \,\right\} .
\ee
Letting $(p,q)$ refer to the signature of $\check C$,
it follows that $\check uq(2;\nu;\tau)$ is equivalent to a representation of $u(p,q;J_0;\tau)$
while $\check hosp(2|2;\nu;\tau)$ is equivalent to a representation of the superalgebra $u(p|q;J_0;\tau)$; the list of isomorphisms is given in Table \ref{Trealform}.

\begin{table}[ht]
\begin{center}
\begin{tabular}{|c|c|c|}
 \hline
 $\nu$ &
 $\check uq(2;\nu;\tau)\cong u(C)$ & $\check hosp(2|2;\nu;\tau)\cong u(C_+|C_-) $
\\\hline\hline
 & & \\[-10pt]
$\nu \geq -1$
&
$u(\infty_++\infty_-)$ & $u(\infty_+|\infty_-)$\\
\hline
 & & \\[-10pt]
\begin{tabular}{l}
$\nu=-(2\ell+1),$\\
$\ell=1,2,...$
\end{tabular}
&
\begin{tabular}{l}
$u(\ell +\frac{ 1+(-1)^\ell }{ 2 },\ell +\frac{ 1-(-1)^\ell }{ 2 } ) $\\$\oplus ~ u(\infty'_++\infty'_-)$\end{tabular}
  &

\begin{tabular}{l}
$u(\ell +\frac{ 1+(-1)^\ell }{ 2 } | \ell +\frac{ 1-(-1)^\ell }{ 2 } )$\\$ \oplus~ u(\infty'_+|\infty'_-)$ \end{tabular}
\\
\hline
 & &\\[-10pt]
\begin{tabular}{l}
$-2\ell-1  > \nu >1-2\ell,$\\
$\ell=1,2,...$
\end{tabular}
&
$\begin{array}{c}
u (\ell , \infty) ,\: \ell=even\\[4pt]
 u (\infty,\ell +1),\: \ell=odd
\end{array}$
&
  $\begin{array}{c}
u (\ell | \infty) ,\: \ell=even\\[4pt]
 u (\infty|\ell +1),\: \ell=odd
\end{array}$
\\
\hline
& & \\[-10pt]
$\nu = -\infty $
&
$u( \infty , \infty)$ &  $u( \infty | \infty)$
\\ \hline
\end{tabular}
\end{center}
\caption{This table displays the $\nu$-dependence of the real forms of the Lie (super)algebras $\check uq(2;\nu;\tau)\cong u(C)$ and $\check hosp(2|2;\nu;\tau)\cong u(C_+|C_-) $. In the above, $u(\eta):=u(p,q)$ if $\eta$ is a diagonal matrix with $p$ positive and $q$ negative entries \emph{idem} $u(\eta_1|\eta_2)$. In the first row, $\infty_\pm$ refer to the dimensions of ${\cal F}_\pm$, and in the second row, $\infty'_\pm$ refer to the dimensions of ${\cal P}_{\infty}{\cal F}_\pm$. The real forms in the graded case (second column) are in agreement with \cite{Bergshoeff:1989ns}. }\label{Trealform}
\end{table}


\section{Chern--Simons formulation}
\label{sec:CS}
\subsection{Blencowe--Vasiliev higher-spin gravity sector}

The Blencowe--Vasiliev higher-spin gravity sector of the fractional-spin gravity model consists of an $lq(2;\nu) \oplus lq(2;\nu)$-valued connection $W$. Its the Fock-space representation reads
\begin{equation}\label{W}
\check W=\frac{1}{4i}\sum_{s=0,1} \Gamma^s \sum_{n\geqslant 0} \sum_{t=0,1}  W_{s,t}^{\,\alpha_1 \cdots \alpha_n}  \, \check k^t  \, \check q_{(\alpha_1 } \cdots \check q_{\alpha_n)}\equiv \frac{1}{4i} \sum_{s=0,1} \Gamma^s \sum_{p,q\geqslant 0}   W_s^{p,q}
 \, |p\rangle \langle
q |\,, \,
\end{equation}
where the gauge-field components in the Fock-space basis are given in terms of those in the multi-spinorial basis via
\begin{equation}\label{WmathC}
W_s^{p,q} =
\sum_{n \geqslant 0}\sum_{t=0,1} W_{s,t}^{\alpha_1
\cdots \alpha_n}(x)\, \mathcal{Q}_{\,\alpha_1 \cdots \alpha_n}{}^{t, p, q }\,,
\end{equation}
using the Fock-space representation matrix of the higher-spin algebra defined by
 \begin{equation}\label{Ycoef}
 \mathcal{Q}_{\alpha_1 \cdots \alpha_n}{}^{t,p,q }:= \langle
 q | \check k^t \, \check q_{(\alpha_1}\cdots
\check  q_{\alpha_n)}  |p
 \rangle \,,
 \end{equation}
which one may think of as a generalized Dirac matrix.

As discussed in the previous section, the connection can be subjected to reality conditions using either complex or hermitian conjugations; for definiteness let us use choose a reality condition of the latter type, namely
\begin{equation}\label{Aconj2}
 \check W^{\check \dagger} = - \check  C \check W \check C \,,
\end{equation}
where $\check C$ is the charge conjugation matrix in \eqref{Yconj} chosen such that \eqref{Creal} and \eqref{Cunimodular} hold. As a result, the multi-spinorial component fields obey
\begin{equation}\label{Aconj1}
(W_{s,t}^{\,\alpha_1 \cdots \alpha_n})^*=(-1)^{nt} \, W_{s,t}^{\,\alpha_1 \cdots \alpha_n}\,  .
\end{equation}
As a consequence of \eqref{C1}, the representation matrix \eqref{Ycoef} obeys
 \begin{equation}\label{ycoefconj}
 \mathcal{Q}_{\, \alpha_1 \cdots \alpha_n}{}^{t, q , l }=  (-1)^{nt}  C_l \, \Big(\mathcal{Q}_{\, \alpha_1 \cdots \alpha_n}{}^{ t, l , q } \Big)^* \, C_q \,.
 \end{equation}
Thus, the master gauge field $W$ obeying \eqref{Aconj2} is represented by a real matrix in the Fock-space basis, \emph{viz.}
\begin{equation}\label{fockreal}
W_s^{p,q}= (W_s^{p,q})^* \, .
\end{equation}

\subsection{Internal color gauge fields}

The fractional-spin gravity also contains an internal color gauge field $U$ given in the
bra-ket basis by
\begin{equation}\label{U}
\check U=\frac{1}{4i}\sum_{s=0,1} \Gamma^s \,\sum_{p,q\geqslant 0}  U^{p}_{s,q} \check T_{p}^{q}\ ,\quad \check T_{p}^{q}:=|q\rangle\langle p|\ .
\end{equation}
It is taken to obey the following reality condition:
\be \check U^{\check \dagger} = - \check U\ ,\ee
such that $\check U$ formally becomes an element of the
$u(\infty)\oplus u(\infty)$ with $u(\infty)$ generated by $\check T_{p}^{q}$,
\begin{equation}
[\check T_{n}^m ,\check T_{q}^{l} ]=i ( \delta^m_q \check T_{n}^{l}- \delta^l_n \check T^m_q ) \,,\quad (\check T^m_n)^{\check \dagger}=\check T^n_m\ .
\end{equation}
With these conventions, it follows that the internal component fields form a hermitian matrix,
\be
(U_{s,p}^q)^\ast= U^p_{s,q}\,.
\ee

\subsection{Hybrid theory with fractional-spin fields}\label{Hybthe}

The higher-spin gravity connection $\check W$ given in \eqref{W} and the internal connection
$\check U$ given in \eqref{U} can be coupled non-trivially via two intertwining one-forms, that we shall
denote by $(\check {\overline{\psi}} , \check {\psi})$, whose gauge symmetries exchange the higher-spin
gravity and internal gauge fields.

In what follows, we present a simplified model exhibiting this feature in which the gauge fields are
further projected using $\Pi_\pm$ as follows:
\begin{eqnarray}
&& \check W_{++}=\check \Pi_+ \check W\, \check \Pi_+=\frac{1}{4i} \sum_{s=0,1}\Gamma^s\sum_{p,q\geqslant 0}   W_s^{2p,2q}
 \, |2p\rangle \langle
2q |\,, \\[5pt]
&& \check U_{--}= \check \Pi_- \check U \,\check \Pi_-=\frac{1}{4i}\sum_{s=0,1}\Gamma^s\sum_{p,q\geqslant 0}  U^{2q+1}_{s,2p+1 } |2p+1\rangle \langle
2q+1 |\,,\\[5pt]
&& \check \psi_{+-} = \check \Pi_+ \check \psi\, \check \Pi_-= \frac{1}{4i} \sum_{s=0,1}\Gamma^s\sum_{p,q\geqslant 0}  \psi^{2p, 2q+1 }_s \, |2p\rangle \langle
2q+1 |\, ,\\[5pt]
&& \check{\overline{\psi}}_{-+} = \check \Pi_- \check {\bar{\psi}}\, \check \Pi_+=\frac{1}{4i} \sum_{s=0,1}\Gamma^s\sum_{p,q\geqslant 0}  \overline{\psi}^{2p}_{s,2q+1}  \, |2q+1\rangle \langle
2p| \, .
\end{eqnarray}
Arranging various master fields into a single two-by-two matrix
\begin{equation}\label{}
\check{\mathbb{A}}=\left[
\begin{array}{cc}
 \check W_{++} & \check \psi _{+-}\\
\check {\overline{\psi}}_{-+}  &  \check U_{--}
\end{array}
\right] \, ,
\end{equation}
the equations of motion can be declared to be of the standard form:
\begin{equation}\label{realF0}
\check{\mathbb{F}}={\rm d} \check{\mathbb{A}} + \check{\mathbb{A}} \wedge \check{\mathbb{A}}=0 \,,
\end{equation}
that is,
\begin{eqnarray}
&& {\rm d} \check W_{++}+ \check W_{++}\w \check W_{++}+\check \psi_{+-}
\w \check {\overline{\psi}}{}_{-+}=0  \, , \\[5pt]
&& {\rm d} \check U_{--}+\check U_{--}\w \check U_{--}+ \check{\overline{\psi}}{}_{-+} \w
\check \psi_{+-} =0  \, , \\[5pt]
&& {\rm d} \check \psi_{+-}+ \check W_{++}\w \check \psi_{+-} + \check \psi_{+-} \w \check U_{--}
=0  \, ,
\\[5pt]
&& {\rm d} \check{\overline{\psi}}{}_{-+} + \check{\overline{\psi}}{}_{-+} \w \check W_{++}
+ \check U_{--} \w \check{\overline{\psi}}{}_{-+}=0  \,,
\end{eqnarray}
which form a non-trivial Cartan integrable system by virtue
of the assignments that we have made so far.
The equations of motion are thus symmetric under the gauge transformations
\begin{equation}
\check{\mathbb{A}} \quad \rightarrow \quad \check{\mathbb{A}}^{\check{\mathbb{G}}} =
\check{\mathbb{G}}^{-1} ({\rm d} + \check{\mathbb{A}} )\, \check{\mathbb{G}} \,, \qquad
\check{\mathbb{G}} = \exp (i\check{\mathbb{X}}) \, ,  \quad
\check{\mathbb{X}} := \left[\begin{array}{cc} \check x_{++}&\check x_{+-}\\\check x_{-+}& \check x_{--} \end{array}\right]\ .
\end{equation}
Thus, $\check W_{++}$ is the connection belonging to the adjoint representation of the non-minimal
bosonic higher-spin subalgebra $lq(2;\nu)_{++}\, \oplus \, lq(2;\nu)_{++}$ of $lq (2;\nu) \, \oplus \,
lq(2;\nu)$;
it consists of all integer spins and has the Fock-space representation
\begin{eqnarray}
\check W_{++}= \frac{1}{4i} \sum_{s=0,1} \Gamma^s \sum_{n\geqslant 0}\sum_{p,q\geqslant}  W_{s,0}^{\alpha_1 \cdots \alpha_{2n}}
 \, \mathcal{Q}_{\alpha_1 \cdots \alpha_{2n}}{}^{0, 2p,2q }\, |2p\rangle \langle
2q | \,.
\end{eqnarray}
The internal gauge field $\check U_{--}$ belongs to the adjoint representation of
$u_{--}(\infty)\oplus u_{--}(\infty)$ where $u_{--}(\infty):=\check\Pi_- u(\infty)\check \Pi_-$.
The intertwining fields $\check \psi_{+-}$ and $\check{\overline{\psi}}{}_{-+} $ belong to
bi-fundamental representations transforming on one side under the higher-spin algebra and on the other side under the internal color gauge algebra.
Thus, the master connection $\mathbb A$ belongs to a hybrid higher-spin algebra, which we refer to as a fractional-spin algebra, consisting of a sector of ordinary higher-spin generators related to space-time symmetries glued to an internal sector of compact generators via a set of intertwining generators belonging to a bi-module.

The action of the global rotation $\check{\mathbb{R}}{}_{2\pi}$ by $2\pi$
generated by $\check x_{++}=2\pi \,   \check \Pi_+ \check J^0$ on the fields is given by
 \begin{eqnarray}
(\check{\mathbb{R}}{}_{2\pi})^{-1} \, \check{\mathbb{A}} \check{\mathbb{R}}{}_{2\pi}
= \left[
\begin{array}{cc}
\check W_{++} & e^{-i \pi \frac{1+\nu}{2}} \check \psi _{+-}\\
e^{i \pi \frac{1+\nu}{2}} \check{\overline{\psi}}{}_{-+}  & \check U_{--}
\end{array}
\right] \,,
\end{eqnarray}
from which it follows that in the semi-classical theory $(\check{\overline{\psi}}, \check \psi)$ have fractional statistical phases $e^{\mp i \pi \frac{1+\nu}{2}}$,  whereas $\check W_{++} $ and $\check  U_{--}$ have bosonic ones.
Thus, the spins and the Grassmann statistics of $(\check{\overline{\psi}}, \check \psi)$ are not correlated in the semi-classical theory for generic values of $\nu$.
Observe that for critical values, $\nu=-2\ell - 1$, the semi-classical statistical phases take the values
\begin{equation}\label{criticalph}
\left\{e^{\mp i \pi \, \ell }: \ell=0,1,2,\dots\right\}= \{ 1,-1,1,-1,\dots \} \, ,
\end{equation}
such that the spins and the Grassmann statistics of $(\check{\overline{\psi}}, \check \psi)$ are correlated in the semi-classical limit for even and odd $\ell$, respectively, in the case of Grassmann even and Grassmann odd fields, in agreement with discussion around \eqref{statisticsq}.

The master connection obeys the following reality condition:
\begin{equation}\label{realityhyb}
\check{\mathbb{A}}^{\check\dagger}  = - \check{\mathbf{C}} \, \check{\mathbb{A}}  \,
\check{\mathbf{C}} \, ,  \qquad \check{\mathbf{C}}
:= \left( \begin{array}{cc} \check C_{++}& 0 \\
0  & \check {\Pi}_{-}
\end{array}\right) \, ,\qquad
\end{equation}
where $\check C_{++}=\check \Pi_+ \check C \check \Pi_+$
whose Fock-space representation is given by
\begin{equation}
\check C_{++}:=\check \Pi_+ \, \check C\,  \check \Pi_+=\sum_{q \geqslant 0} C_{2q} \, |2q\rangle \langle
2q | \, .
\end{equation}
and
$\check{\Pi}_-=\sum_{q \geq 0}   |2q+1\rangle \langle 2q+1 |$.

As discussed in Section \ref{Sec:bases}, the key issue is the choice of bases used to expand the various
gauge fields.
Strictly speaking, the fractional-spin gravity model for which we have an off-shell formulation, is based
on a master field
\begin{equation}\label{}
\mathbb{A} \in   \left[
\begin{array}{cc}
lq_{++}(2;\nu; \mathbb{R})  & \rhoF\left({\rm Bi}(lq_{++}(2;\nu; \mathbb{R})|u_{--}(\infty))\right)\\[5pt]
\rhoF\left(\overline{{\rm Bi}}(u_{--}(\infty)|lq_{++}(2;\nu; \mathbb{R}))\right) & \rhoF\left(u_{--}(\infty )\right)
\end{array}
\right]\otimes {\rm Cliff}_1(\Gamma) \, ,
\end{equation}
where $\rhoF$ denotes a morphism from ${\rm End}({\cal F})$ to the oscillator
algebra and ${\rm Bi}(a|b)$ denotes a bi-module with a left action of $a$ and a right action of $b$.
The higher-spin connection is thus expanded in multi-spinorial basis, in which only the trivial element has a diagonal one-sided action, while the internal gauge field and the intertwiners are expanded in the Fock-space basis, in which the spin operator $\check{J}^0$ has diagonal one-sided actions.
The role of the map $\rhoF$ is to realize the latter basis elements as elements of the oscillator algebra rather than ${\rm End}({\cal F})$ as to make sense of the source term
$\psi\overline{\psi}$ in the equation for $W_{++}$.

As far as the on-shell formulation is concerned, it follows from Eqs. \eqref{gl1} and \eqref{gl2}, that undoing of the map $\rhoF$ by mapping $\mathbb A$ to its representation $\check{\mathbb{A}}$ in ${\cal F}$ yields a model that is equivalent to the original one only for non-critical $\nu$.
However, as outlined in Section \ref{Sec:main}, the preference for the former model, formulated in terms of $\mathbb{A}$ rather than $\check{\mathbb{A}}$, stems from the fact that the construction of the standard Chern--Simons action \eqref{CS1} requires the introduction of a bi-linear form \eqref{traceoperation} on the fractional-spin algebra.
This bi-linear form is based on the trace operation \eqref{Trace} in its turn based on the supertrace operation \eqref{STr} whose implementation is straightforward once all objects have been mapped to the star-product algebra\footnote{Alternatively, it would be interesting to investigate whether it is possible to start from an implementation of the supertrace operation in ${\rm End}({\cal F})$ and seek a scheme for regularizing the supertraces of the multi-spinorial generators of $lq(2;\nu)$.}

\subsection{Finite-dimensional truncations at critical $\nu$}
\label{ftrunc}

In account of the discussion surrounding Eq. \eqref{caseIII}, for critical values of $\nu=- 2 \ell - 1$, the algebra $Aq(2;\nu)\oplus Aq(2;\nu)$  possesses an additional decomposable structure in finite and infinite dimensional subsectors, so that in those cases the connection splits into
\begin{equation}
\check{\mathbb{A}}= \check{\mathbb{A}}_{\rm f}+ \check{\mathbb{A}}_\infty \, , \qquad  \check{\mathbb{A}}_{\rm f}:= \mathcal{P}_{\rm f} \,  \check{\mathbb{A}} \, \check{\mathcal{P}}_{\rm f}  \, , \quad \check{\mathbb{A}}_\infty := \mathcal{P}_\infty \,  \check{\mathbb{A}} \, \mathcal{P}_\infty \, ,
\end{equation}
where, in the notation of Eq. \eqref{gl1}, one has
\be  \check{\mathbb{A}}_{\rm f} \, \in \,  gl(2\ell+1) \, \oplus \,  gl(2\ell+1)\, , \quad \check{\mathbb{A}}_\infty \, \in \,  gl(-\ell;J_0;0)\, \oplus \, gl(-\ell;J_0;0)\ .\label{AfAinf}\ee
With our election of the representation of the oscillators generators \eqref{a+a-}, other projections vanish, \emph{i.e.}
\begin{equation}
\mathcal{P}_{ {\rm f} \atop \infty } \check{\mathbb{A}} \, \mathcal{P}_{ \infty \atop {\rm f}  } = 0 \, .
\end{equation}
The basis element of the subsectors $\mathbb{A}_{\rm f}$ and $\mathbb{A}_{ \infty}$ are given respectively by
\begin{eqnarray}
gl(2\ell+1)|_{2\ell+1} \, \oplus \,  gl(2\ell+1)|_{2\ell+1}  & = & \{ \Gamma^s \, k^t \, \mathcal{P}_{\rm f}\, q_{(\alpha_1}\cdots  q_{\alpha_n)} \, \mathcal{P}_{\rm  f}\,, : \, s,t=0,1;\,n=0,...,2\ell\,  \} \,, \nonumber\\[5pt]
Aq(2;2\ell+1)|_{{\cal F}} \, \oplus \, Aq(2;2\ell+1)|_{{\cal F}} & =  &\{ \Gamma^s \, k^t \, \mathcal{P}_{\infty}\,  q_{(\alpha_1}\cdots  q_{\alpha_n)} \, \mathcal{P}_{\infty}\,, :\, s,t=0,1;\, n=0,... \,  \} \,. \nonumber
\end{eqnarray}
One can verify that the associative algebra spanned by
$ k^t \, \mathcal{P}_{\rm f}\,\, q_{(\alpha_1}\cdots  q_{\alpha_n)} \, \mathcal{P}_{\rm f}$ with $n=0,...,2\ell \,$; $t=0,1\,$, is isomorphic \cite{Vasiliev:1989re}
to Mat${}_{2\ell+1},(\mathbb{C})$
by counting the number of independent generators  considering the identity
\begin{equation}
 \mathcal{P}_{\rm f}\, q_{(\alpha_1}\cdots  q_{\alpha_{2\ell})} \, \mathcal{P}_{\rm  f} \equiv
 \, k \, \mathcal{P}_{\rm f}\, q_{(\alpha_1}\cdots  q_{\alpha_{2\ell})} \, \mathcal{P}_{\rm  f} \quad \Leftrightarrow\quad \Pi_- \, \mathcal{P}_{\rm f}\, q_{(\alpha_1}\cdots  q_{\alpha_{2\ell})} \, \mathcal{P}_{\rm  f}  =0 \, . \end{equation}
In this way, the hybrid model \ref{Hybthe} constructed in Fock space, including the correspondent reality conditions \eqref{realityhyb}, thus decomposes into a finite-dimensional and an infinite-dimensional model\footnote{Thus, these equations for $\mathbb{A}$ decompose in a different way, such that $\mathbb{A}_{\rm f}$ sources $\mathbb{A}_{\infty}$.}
\begin{equation}\label{}
\check{\mathbb{F}}_{\rm f}=d\check{\mathbb{A}}_{\rm f}+ \check{\mathbb{A}}_{\rm f}\w \check{\mathbb{A}}_{\rm f}=0 \,, \qquad \check{\mathbb{F}}_\infty=d \check{\mathbb{A}}_\infty+ \check{\mathbb{A}}_\infty\w \check{\mathbb{A}}_\infty =0 \, , \qquad  \check{\mathbb{A}}_{\rm f}\w \check{\mathbb{A}}_\infty=\check{\mathbb{A}}_\infty \w \check{\mathbb{A}}_{\rm f}=0 \, ,
\end{equation}
for the corresponding algebras \eqref{AfAinf}; that is
\begin{equation}
\check{\mathbb{A}}_{\rm f} \quad \in \quad  \left[
\begin{array}{cc}
gl(\ell+1, \mathbb{R})   & {\rm Bi}(\ell+1\otimes \overline\ell) \\
\overline{{\rm Bi}}( \ell\otimes \ell+1) & u(\ell)
\end{array}
\right]\otimes {\rm Cliff}_1(\Gamma) \,,
\end{equation}
and
\begin{equation}
\check{\mathbb{A}}_{ \infty} \quad \in \quad  \left[
\begin{array}{cc}
gl(-\ell;J_0;0)   & {\rm Bi}((-\ell;J_0;0)\otimes \overline\infty) \\
\overline{\rm Bi}(\infty\otimes (-\ell;J_0;0)) & u(\infty)
\end{array}
\right]\otimes {\rm Cliff}_1(\Gamma) \,,
\end{equation}
where ${\rm Bi}(v\otimes w)$ denotes a bi-module consisting of a left-module $v$ and a right-module $w$.

\subsection{Truncations of color gauge fields}

To begin with, for any $\nu$ and $N\in \mathbb{N}$, it is possible to choose $u(N)$ subalgebras of ${\rm End}({\cal F})$ and truncate $U\in u(N)$ and simultaneously take $\psi$ and $\overline\psi$, respectively, to transform in $\bar N$ and $N$.
For any given $N$, there exists an infinite number of such level truncations. 

Another type of truncation of the color gauge fields is possible in the non-unitary regime $\nu<-1$. Here one notes that if $\check \psi= | \sigma \rangle\langle c |$, where thus $\sigma$
is a spin and $c$ is a color, then
$\check{\overline{\psi}} = - | c \rangle\langle \sigma | C$ and hence
$\check{\psi} \check{\overline{\psi}} = |\sigma \rangle\langle \sigma | C$ while
$\check{\overline{\psi}} \check  \psi = |\sigma\rangle\langle c|C|c\rangle\langle \sigma|$ that
can vanish in the non-unitary regime.
Thus, the fractional-spin fields necessarily source the tensor-spinorial higher-spin gravity field $W$
(\emph{c.f.} positivity of energy in ordinary gravity) 
while the internal gauge field $\check U$ can be truncated consistently leading to
\begin{equation}
{\rm d}\check W+\check W^2 +\check \psi \, \check{\overline{\psi}}=0\ ,\quad
{\rm d}\check \psi +\check W \check \psi=0\ ,\quad
{\rm d}\check{\overline{\psi}} + \check{\overline{\psi}} \,\check W=0\ ,
\quad \check{\overline{\psi}}\,\check \psi=0\ ,
\end{equation}
which defines a quasi-free differential algebra. Thus, the last constraint above possesses non-trivial solutions
owing the non-definite signature of the invariant conjugation matrix of the  representation of the higher
spin algebra carried by the fractional spin  fields, while $\check  \psi\,  \check{\overline{\psi}}=0$
does not have non-trivial solutions.

\section{Conclusions}
\label{sec:Ccls}

In this paper, we have presented a new class of
three-dimensional Chern--Simons higher-spin gravities 
that we refer to as fractional-spin gravities.
These theories are extensions of ordinary 
Blencowe--Vasiliev  \cite{Blencowe:1988gj,Vasiliev:1989re}
higher-spin gravities and Chern--Simons
gauge theories by bi-fundamental one-forms 
valued in direct products of fundamental
representations the higher-spin algebras 
and the internal compact gauge algebras.
In effect, the fractional-spin 
models have been obtained by a non-standard
embedding of the Lorentz algebra into an 
original enlarged Blencowe--Vasiliev model; 
in this sense one may interpret the fractional-spin
gravities as describing new vacuum sectors of
the Blencowe--Vasiliev theory, as we shall 
comment on more below.

The fundamental representations of
the higher-spin algebras are infinite-dimensional and 
characterized by a deformation parameter 
$\nu\in\mathbb{R}$: For non-critical $\nu$
they remain irreducible under the Lorentz 
sub-algebra with spin $\frac14(1+\nu)$; for 
critical $\nu=-1,-3,\dots$ they decompose
into a finite-dimensional tensor or tensor-spinor
and an infinite-dimensional representation
with spin $-\nu$.
The color indices, on the other hand,
can be chosen to be finite-dimensional
by level truncation, and if the fractional-spin
representation is non-unitary, that is, if $\nu<-1$
then the internal gauge fields can be truncated;
the theory then consists only of the higher-spin
gravity fields and the fractional-spin fields.

Denoting the Blencowe--Vasiliev connection by
$W$, which thus consists of a collection of
Lorentz-tensorial gauge fields making up the 
adjoint representation of the higher-spin algebra, 
and the fractional-spin fields and internal connection 
by $(\psi,\overline\psi)$ and $U$, respectively, 
we have proposed to describe the fractional-spin gravities
on-shell using the following integrable system of equations:
\begin{eqnarray} {\rm d}  W+{W}{}^2+
  \psi {} {\overline\psi} = 0\ ,\quad
{\rm d}  \psi +  W {}  \psi+  \psi {}  U=0\ ,\end{eqnarray}
\begin{eqnarray} {\rm d}{\overline{\psi}}
+{\overline{\psi}} {}  W + U {} {\overline{\psi}} = 0\ ,\quad
{\rm d}  U+ U{}  U +{\overline{\psi}}{}  \psi=0\ ,\end{eqnarray}
or more concisely, as 
\begin{eqnarray} {\rm d} {\mathbb{A}} +
{\mathbb{A}}{} {\mathbb{A}} =0\ ,
\quad {\mathbb{A}} =\left[\begin{array}{cc}  W &
 \psi\\ {\overline{\psi}} &
 U \end{array}
\right]\ .\end{eqnarray}
The underlying fractional-spin algebra carries a 
$\mathbb{Z}_2$-grading similar to that of ordinary 
superalgebras: The fractional-spin generators close 
onto higher-spin and internal generators,
while the higher-spin and internal generators rotate 
the fractional-spin charges into themselves.
Thus, the fractional-spin fields transform under 
one-sided actions of the higher-spin and internal
Lie, and the fractional-spin transformations can 
send higher-spin gauge fields into internal gauge 
fields and vice versa. 

We would like to stress that the simple appearance 
of the construction is due to the fact that it relies 
on the consistent fusion of two sectors of the
enveloping algebra of the Wigner--Heisenberg deformed
oscillators: The sector of arbitrary polynomials 
in deformed oscillators can be combined with the 
sector of Fock-space endomorphisms into an associative
algebra by realizing the latter as elements of the
enveloping algebra. 
In this paper, we have demonstrated this algebraic
structure at the level of Fock-space representations, 
which are sufficient for the on-shell formulation.
The off-shell formulation requires, however, the 
implementation using enveloping algebra techniques, 
as to realize the bi-linear form going into the 
definition of the Chern--Simons action; we leave 
a more detailed description of the off-shell 
formulation as well as the construction 
of non-topological fractional-spin models
for forthcoming works.

In terms of $sl(2)$ representation theory, the fractional-spin representations belong to the discrete series \cite{Bargmann:1946me} which are lowest-weight representations in the compact basis, labeled by the lowest eigenvalue of the spatial rotation generator $J^0$ of $so(2,1)\cong sl(2)$.
Generic values of the lowest spin imply irreducibility, while negative integer or negative half-integer lowest spins, respectively, imply decomposability with finite-dimensional invariant tensor or tensor-spinorial subspaces.
Hence, finite-dimensional higher-spin models can be singled out; by combining various reality conditions and working with fractional-spin fields that are either bosons or fermions one may arrive at models based on $sl(N)$, $su(p,q)$ or $su(p|q)$.

The fact that the fractional-spin fields $(\psi,\overline\psi)$ 
are constructed from tensor-spinor higher-spin fields by a change 
of basis, can be interpreted as that the latter condense
into the former in a new vacuum of the Prokushkin-Vasiliev 
system where color interactions emerge. 
This phenomena is reminiscent of how new phases can
be reached in strongly correlated systems by means of
large gauge transformation, as for example in the confined 
phase of QCD according to t'Hooft's mechanism \cite{'tHooft:1977hy}.
It is thus inspiring to entertain the idea that 
the new vacua of Blencowe--Vasiliev theory studied 
arise in a similar fashion, namely, via a large gauge 
transformation of the Blencowe--Vasiliev vacuum formed 
by tensor and tensor-spinor fields.   
This physical picture also resembles the fractional 
quantum Hall effect \cite{Wilczek:1982wy,Laughlin:1983fy,Halperin:1984fn,
dePicciotto:1997qc,Polychronakos:2001mi}
where many-electron systems exposed to strong magnetic fields  
become confined giving rise to quasi-particle anyons.

As mentioned already, anyons can be obtained
 in the form of a Wilson line coming from infinite and attached in its extreme to a charged particle \cite{Itzhaki:2002rc}, yielding the transmutation to braided statics. Although we have not discussed these aspects in this paper, it suggests by analogy  that we may be
in a similar picture, namely that the fractional spin fields should correspond to Wilson lines attached to
the AdS boundary and to some higher spin particles
with usual boson or fermion statistics, although in
the present states of our theory the latter particles
must also be located at the boundary, as it happens in
Chern--Simons theory where the dynamical degrees of
freedom are confined to the boundary.

It is worth to mention that open higher-spin Wilson-lines have been analyzed recently
\cite{Ammon:2013hba,deBoer:2013vca} and their insertions have been argued to be dual to sending the dual conformal
field theory to phases with finite entanglement entropies.
One problem that one can investigate, starting
from our model, is a particular type of classical solutions in fractional-spin
gravity that may have an interpretation as entanglement entropy.
Along the same lines, a suitable approach could be suggested by the considerations
made in the work \cite{Compere:2013nba}.

\section*{Acknowledge}

We thank Fabien Buisseret, Andrea Campoleoni, Alejandra Castro, 
Johan Engquist, Matthias Gaberdiel, Dileep Jatkar, Soo-Jung Rey,
Kostas Siampos and Philippe Spindel for discussions.
The work of N.B. was supported in part by
an ARC contract n$^{\rm o}$ AUWB-2010-10/15-UMONS-1.
M. V. is supported by FONDECYT postdoctoral grant n$^{\rm o}$ 3120103.
P.S. and M.V. acknowledge the support of the F.R.S.-FNRS ``Ulysse'' Incentive Grant for Mobility in Scientific Research during the first stage of this work, which was carried out at the Service de M\'ecanique et Gravitation at UMONS.

\providecommand{\href}[2]{#2}\begingroup\raggedright\endgroup

\end{document}